\documentclass[aps,prl,twocolumn,twoside,amsmath,raggedbottom,superscriptaddress]{revtex4-1}

\usepackage{txfonts}

\usepackage[english]{babel}

\usepackage{amsmath}
\usepackage{amssymb}
\usepackage{amsfonts}

\usepackage{graphicx}

\newcommand{\LO}{\ensuremath{\text{LO}}}
\newcommand{\NLO}{\ensuremath{\text{NLO}}}
\newcommand{\NNLO}{\ensuremath{\text{N}^2\text{LO}}}
\newcommand{\NNNLO}{\ensuremath{\text{N}^3\text{LO}}}

\newcommand{\cD}{\ensuremath{c_{\text{D}}}}
\newcommand{\cE}{\ensuremath{c_{\text{E}}}}

\begin{document}

\title{Family of Chiral Two- plus Three-Nucleon Interactions for Accurate Nuclear Structure Studies}

\author{Thomas~H\"uther}
\email[]{huether@theorie.ikp.physik.tu-darmstadt.de}
\affiliation{Institut f\"ur Kernphysik, Technische Universit\"at Darmstadt, 64289 Darmstadt, Germany}

\author{Klaus~Vobig}
\affiliation{Institut f\"ur Kernphysik, Technische Universit\"at Darmstadt, 64289 Darmstadt, Germany}

\author{Kai~Hebeler}
\affiliation{Institut f\"ur Kernphysik, Technische Universit\"at Darmstadt, 64289 Darmstadt, Germany}
\affiliation{ExtreMe Matter Institute EMMI, GSI Helmholtzzentrum f\"ur Schwerionenforschung GmbH, 64291 Darmstadt, Germany}

\author{Ruprecht~Machleidt}
\affiliation{Department of Physics, University of Idaho, Moscow, Idaho 83844, USA}

\author{Robert~Roth}
\email[]{robert.roth@physik.tu-darmstadt.de}
\affiliation{Institut f\"ur Kernphysik, Technische Universit\"at Darmstadt, 64289 Darmstadt, Germany}

\date{\today}

\begin{abstract}

We present a family of nucleon-nucleon (NN) plus three-nucleon (3N) interactions up to \NNNLO{} in the chiral expansion that provides an accurate \emph{ab initio} description of ground-state energies and charge radii up to the medium-mass regime with quantified theory uncertainties. Starting from the NN interactions proposed by Entem, Machleidt and Nosyk, we construct 3N interactions with consistent chiral order, non-local regulator, and cutoff value and explore the dependence of nuclear observables over a range of mass numbers on the 3N low-energy constants. By fixing these constants using the $^{3}$H and $^{16}$O ground-state energies, we obtain interactions that robustly reproduce experimental energies and radii for large range from p-shell nuclei to the nickel isotopic chain and resolve many of the deficiencies of previous interactions. Based on the order-by-order convergence and the cutoff dependence of nuclear observables, we assess the uncertainties due the interaction, which yield a significant contribution to the total theory uncertainty.  

\end{abstract}

\maketitle

\paragraph*{Introduction.} 
Chiral effective field theory (EFT) has evolved into the standard approach for the construction of nuclear interactions over the past decade. A large variety of nuclear structure and reaction calculations, ranging from light systems up to mass numbers of 100 and beyond, have been performed starting from chiral EFT interactions---some recent highlights are presented in Refs.~\cite{TaSa19,MaHu19,GyHa19,MoSi18,GaBi16}. For a long time, many-body practitioners employed a few specific chiral Hamiltonians, often based on the same chiral NN interaction at \NNNLO{} by Entem and Machleidt \cite{EnMa03}, supplemented with a 3N force at \NNLO{} with a local regulator \cite{Na07,GaQu09,GaQu19}. Interactions with non-local regulators at different chiral orders were available \cite{EpNo02,EpGl05}, but rarely used in nuclear structure applications beyond the few-body sector \cite{CaRo16}. Although these interactions were successful in many applications, they do exhibit systematic deficiencies, most significantly, a systematic underestimation of nuclear radii, particularly in the regime of medium-mass nuclei starting with the oxygen isotopic chain \cite{LaSo16,BiLa14}. 

This prompted the development of interactions that included many-body observables into the construction of the interaction. One example is the $\NNLO_{\text{SAT}}$ interaction \cite{EkHa15} that uses additional information on oxygen ground-state energies and radii in a combined fit of an NN+3N interaction at \NNLO{}. However, the improved description of ground-state observables comes at the price of a degraded reproduction of phase shifts, which affects the spectroscopy of light and medium-mass nuclei. Other variants of interactions with improved binding energies have been proposed \cite{HeBo11,SoNa19}, which, however, do not resolve the problem with radii. Besides the practical deficiencies, none of these interaction capitalizes on the systematic character of chiral EFT. The dependence of nuclear observables on the order of the chiral expansion provides a direct approach to the quantification of theory uncertainties. This order-by-order uncertainty analysis was established in few-body systems \cite{EpKr15,BiCa16,BiCa18} and recently applied in nuclear-matter calculations up to \NNNLO{} \cite{DrHe19}. First applications of this uncertainty analysis for energies of light and medium-mass nuclei have been disucssed in \cite{BiCa18} with NN interactions and in \cite{EpGo19} with NN+3N interactions up to \NNLO{}.
  
In this Letter we present a new family of chiral NN+3N interactions that addresses the aforementioned issues. Starting from a set of chiral NN interactions by Entem, Machleidt and Nosyk (EMN) for a range of chiral orders and cutoffs \cite{EnMa17}, we construct 3N interactions using the same chiral order, the same regulator scheme, and the same regulator scale as in the NN sector (as also done in Ref. \cite{HoDr19}). We explore the dependence of few and many-body observables on the low-energy constants (LEC) in the 3N sector. Using the $^{16}$O ground-state energy to constrain $c_\text{D}$ and the triton ground-state energy to fix $c_\text{E}$, we establish a family of NN+3N interactions for all orders up to \NNNLO{} with cutoffs 450, 500, and 550 MeV. We employ no-core shell model (NCSM), in-medium similarity renormalization group (IM-SRG), and in-medium no-core shell model (IM-NCSM) calculations to explore a range of observables up into the nickel isotopic chain. We exploit the order-by-order and cutoff systematics to quantify the uncertainties of all many-body observables, including the uncertainties of the many-body scheme. The accuracy of this new family of interactions is remarkable and will certainly prompt a range of future studies.

\paragraph*{Computational Framework.}
For a systematic investigation of nuclear observables and their dependence on the underlying chiral NN+3N interaction, we need efficient many-body methods with controlled uncertainties. Since we address a broad mass range and a variety of observables, from ground-state energies and radii to excitation spectra, we will adopt different state-of-the-art ab initio methods. 
 
For the description of few-body systems we use a Jacobi harmonic-oscillator formulation of the NCSM with the bare Hamiltonian, which allows us to access large model spaces with little computational effort and small uncertainties. For $^3$H we use model spaces up to $N_{\max}=48$ with frequencies $\hbar\Omega\approx 20-32$\,MeV around the variational energy minimum, which warrants energy convergence to better than $1$keV. For the calculation of $^4$He ground-state observables we use $N_{\max}=24$ and frequencies $\hbar\Omega\approx 20-32$\,MeV, which leads to typical many-body uncertainties of $10$\,keV and $0.001$\,fm for energies and radii, respectively. 
 
For surveys of ground-state energies and radii of closed-shell nuclei up into the nickel isotopic chain, we employ the single-reference IM-SRG \cite{He16,HeBo16,HeBo13,TsBo11} in a Magnus formulation, truncated beyond normal-ordered two-body terms, for an efficient calculation of radii. We use a consistent free-space SRG evolution of the Hamiltonian (up to three-body terms) and the radius operator (up to two-body terms) with a typical flow parameter $\alpha=0.04\,\text{fm}^4$, corresponding to a momentum scale of $2.24\,\text{fm}^{-1}$ \cite{RoLa11,RoCa14}. In addition, we use the natural-orbital single-particle basis extracted for a perturbatively corrected one-body density matrix of the target nucleus \cite{TiMu19}. 
   
For the description of open-shell nuclei we employ the IM-NCSM introduced in Ref.~\cite{GeVo17}. It is based on a multi-reference IM-SRG evolution of the Hamiltonian and all other operators of interest, starting from a multi-configurational reference state from an NCSM calculation in a small reference space, typically $N_{\max}^{\text{ref}}=0$ or $2$. This evolution suppresses pieces of the $A$-body Hamiltonian that couple the reference space to the rest of the model space, thus, leading to an extremely fast convergence of a subsequent NCSM calculation. As for the single-reference IM-SRG, we employ a free-space SRG evolution and a natural-orbital basis. For light p-shell nuclei we also show conventional NCSM calculations with the harmonic-oscillator basis.

\begin{figure}
\includegraphics[width=1\columnwidth]{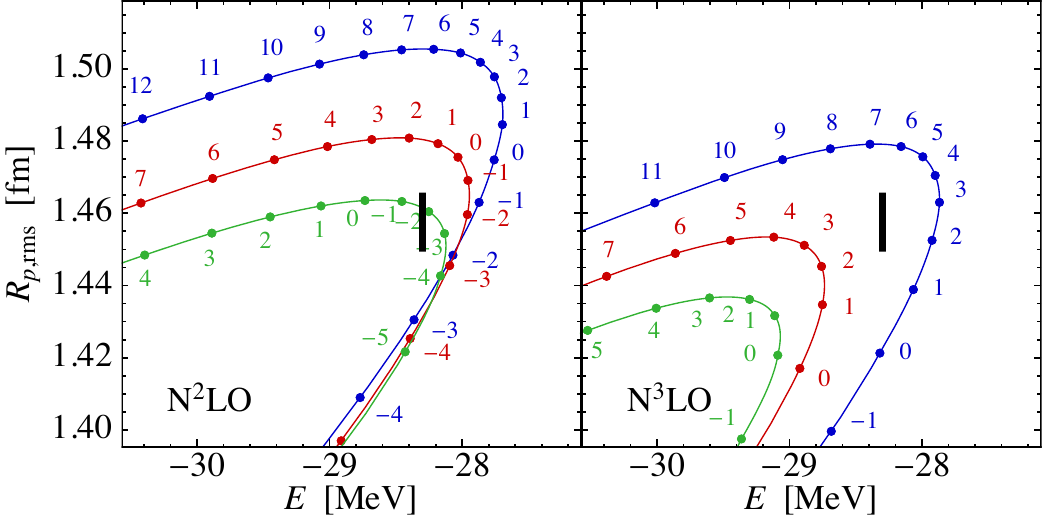}
\caption{Ground-state energy and rms-radius of $^4$He as parametric function of the low-energy constant \cD{} (see labels) for NN+3N interactions at \NNLO{} (left) and \NNNLO{} (right) for cutoffs $\Lambda=450$\,MeV (blue), $500$\,MeV (red), $550$\,MeV (green). For each \cD{}, the corresponding \cE{} is determined to reproduce the $^3$H ground-state energy. \label{fig:he4_cDvar}}
\end{figure}

\paragraph*{New Family of Non-Local NN+3N Interactions.}

In a first step towards the construction of a family of non-local NN+3N interactions up to \NNNLO{}, we consider the few-nucleon systems $^3$H and $^4$He. We employ the EMN interactions from LO to \NNNLO{} with non-local regulators and cutoffs $\Lambda=450, 500,$ and $550$\,MeV. They are supplemented with the corresponding 3N interactions at \NNLO{} and \NNNLO{}, computed using the framework of Ref. \cite{HeKr15}, with non-local regulators in the Jacobi momenta $p$ and $q$ of the form $\exp(-((p^2 + 3/4q^2)/\Lambda^2)^{n})$ with the same $\Lambda$ as in the NN interaction. We will adopt $n=3$ in the following---choosing another value will lead to slight shifts in the values of the LEC, but will not change many-body results significantly. Unlike many previous studies, we do not fix \cD{} in the few-body domain, e.g., by using the triton half-life or the $^4$He radius. We keep \cD{} as a free parameter and only fix \cE{} for a range of different \cD{} by fitting the triton ground-state energy. In this way, we can study different many-body observables and their dependence on \cD{}, before deciding on a selection criterion for the optimum \cD{}.

\paragraph*{Exploring \cD{} in Few-Body Systems.}

As a first set of observables for this analysis, we consider the ground-state energy $E$ and point-proton root-mean-square (rms) radius $R_{p,rms}$ of $^4$He obtained in NCSM calculations with the bare NN+3N interactions at \NNLO{} and \NNNLO{}. In Fig.~\ref{fig:he4_cDvar} we present the results in form of \cD{}-trajectories in the $(E,R_{p,rms})$-plane for the three different cutoffs. All \cD{} trajectories follow rotated parabolic curves, which shift systematically to lower energies and radii with increasing cutoff. There is an upper bound for the $^4$He ground-state energy and in some cases, e.g., for the \NNNLO\ interaction at $\Lambda=500$\,MeV, this makes it impossible to reproduce the experimental ground-state energy---for all \cD{} $^4$He is overbound. Another interesting implication relates to the Tjon-line, i.e., the correlation between the $^3$H and $^4$He ground-state energies \cite{PlHa05,NoKa00}. For all interactions and all \cD{} values used here, the $^3$H ground-state energy is fixed to the experimental value through fitting \cE{}. Nevertheless, the \cD{} variation changes the $^4$He energy over a substantial range, thus, departing from the Tjon-line in a systematic way.

\begin{figure}
\includegraphics[width=0.9\columnwidth]{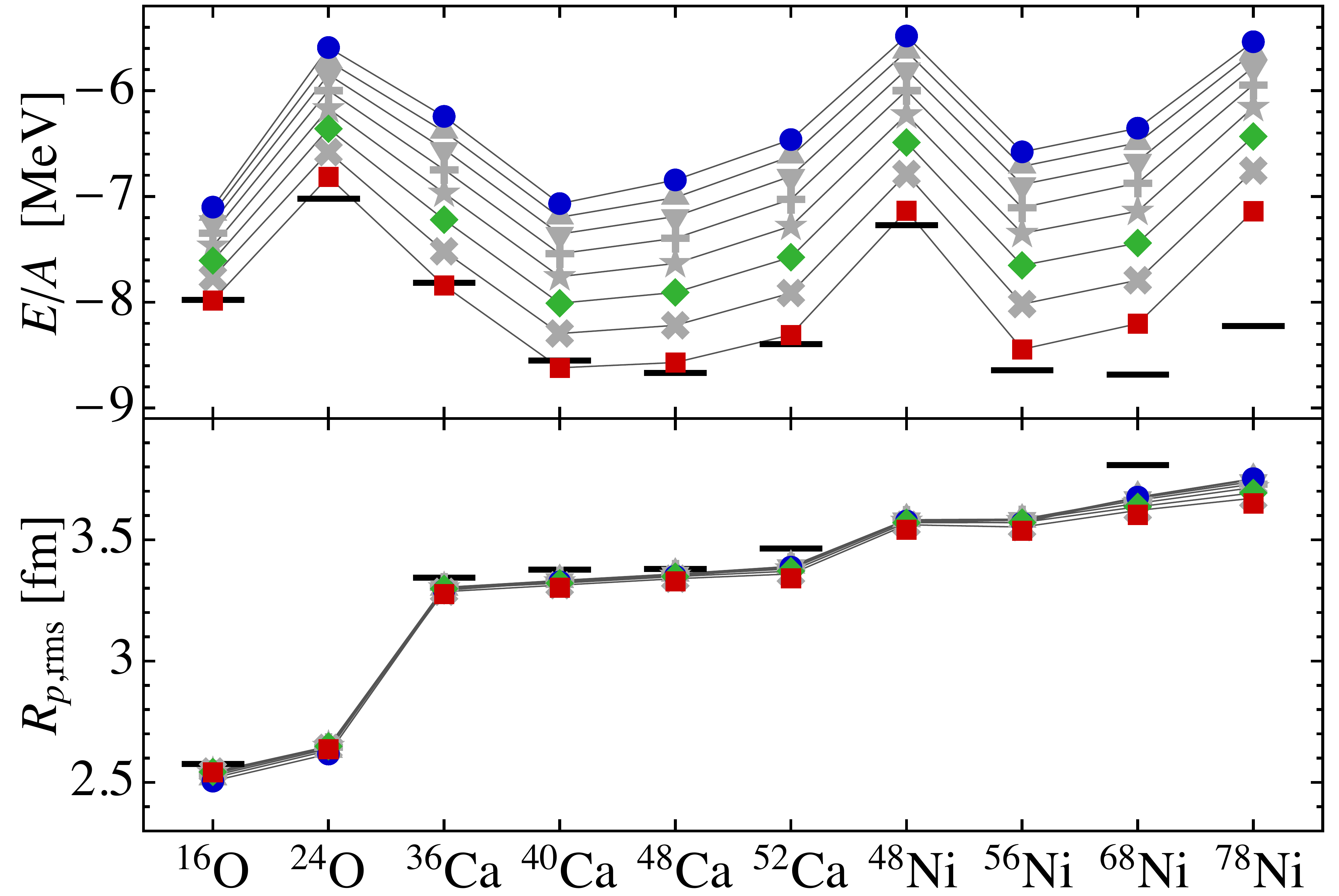}
\caption{Ground-state energies and point-proton rms radii for selected medium-mass isotopes obtained in IM-SRG for NN+3N interaction at \NNNLO{} with $\Lambda=500$\,MeV for a range of \cD{} parameters from $-3$ (blue) to $+4$ (red) in steps of $1$. \label{fig:medmass_cDvar}}
\end{figure}

\begin{table}
\caption{Values of the 3N low-energy constants obtained from considering the $^{16}$O ground-state energy. 
The values for the two-pion LECs are $(c_1,c_3,c_4)=(-0.74,-3.61,2.44)\,\text{GeV}^{-1}$ for \NNLO{}, $(-1.20,-4.43,2.67)\,\text{GeV}^{-1}$ for \NNNLO{}, $(-1.07,-5.32,3.56)\,\text{GeV}^{-1}$ for \NNNLO{}', taken from \cite{EnMa17}. We use isospin-averaged values for $C_{\text{S}}=(-4.60,-4.78,-4.56)\,\text{fm}^2$ and $C_{\text{T}}=(-0.010,-0.163,-0.069)\,\text{fm}^2$ for the three cutoffs $(450, 500, 550)\,\text{MeV}$.}
\label{tab:lecs}
\begin{ruledtabular}
\begin{tabular}{l c r r c c }
 & $\Lambda$ [MeV] & \cD{} & \cE{} & $E$($^4$He) [MeV] & $R_{\text{rms}}$($^4$He) [fm] \\
\hline
\NNLO{}   & 450  & 10.0 &   0.909 & -29.46 & 1.498\\
\NNNLO{}  & 450  &  9.0 &  -0.152 & -29.05 & 1.475\\
\NNNLO{}' & 450  &  9.0 &   0.544 & -29.50 & 1.499\\
\hline
\NNLO{}   & 500  &  5.0 &  -0.159 & -29.42 & 1.475\\
\NNNLO{}  & 500  &  4.0 &  -1.492 & -29.12 & 1.453\\
\NNNLO{}' & 500  &  4.0 &  -1.481 & -29.41 & 1.497\\
\hline
\NNLO{}   & 550  &  2.0 &  -0.966 & -29.45 & 1.459\\
\NNNLO{}  & 550  &  3.0 &  -1.745 & -29.60 & 1.437\\
\NNNLO{}' & 550  &  1.0 &  -3.412 & -29.64 & 1.477\\
\end{tabular}
\end{ruledtabular}
\end{table}

\begin{figure*}
\includegraphics[width=\textwidth]{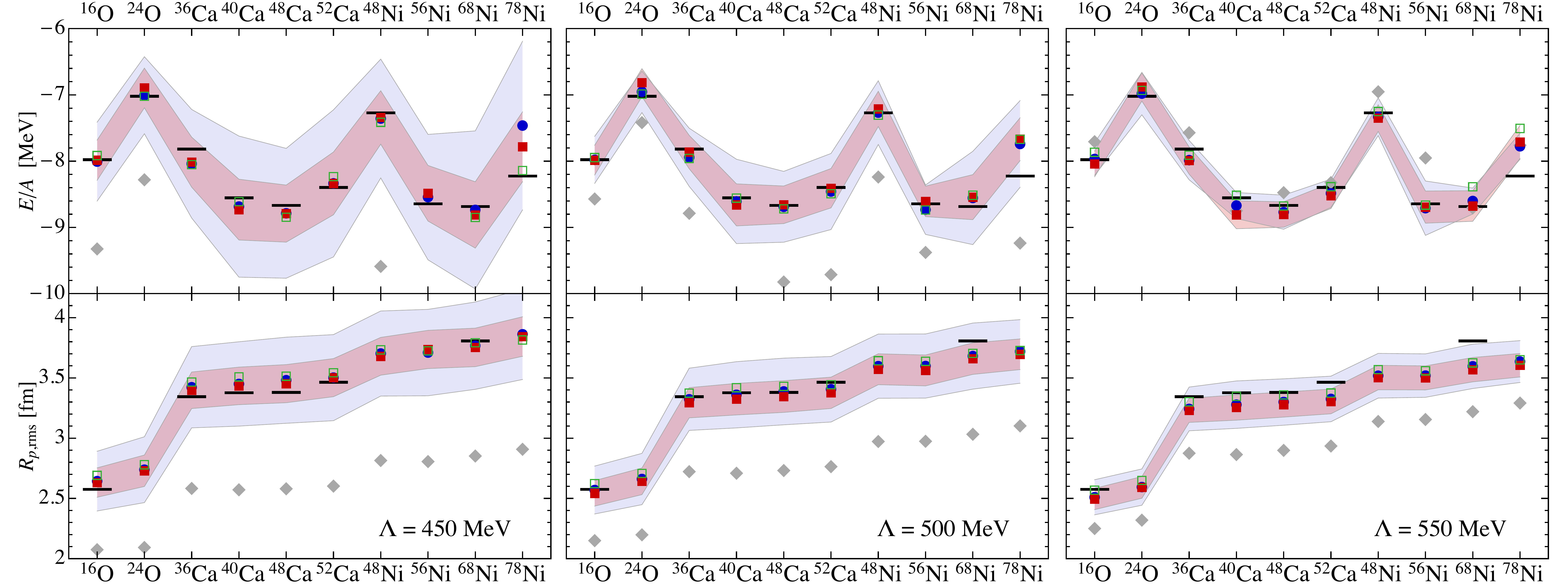}
\caption{Ground-state energies (top panels) and point-proton rms radii (bottom panels) obtained in IM-SRG calculations for the \NLO{} (solid gray diamonds), \NNLO{} (blue circles), \NNNLO{} (red boxes), and \NNNLO{}' (open green boxes) interactions with $\Lambda=450$\,MeV (left), $500$\,MeV (center), and $550$\,MeV (right). The error bands for \NNLO{} (blue) and \NNNLO{} (red) are derived from the order-by-order behavior and include the many-body uncertainties (see text). Experimental data is indicated by black bars \cite{AME2016,AnMa13,GaBi16}.\label{fig:medmass}}
\end{figure*}

\paragraph*{Exploring \cD{} in Many-Body Systems.}

We can repeat this analysis for ground-state energies and point-proton radii of heavier nuclei, ranging from the oxygen to the nickel isotopic chain. For simplicity we limit ourselves to selected closed-shell isotopes and use single-reference IM-SRG calculations. The results for a variation of \cD{} at \NNNLO{} with $\Lambda=500$\,MeV are presented in Fig.~\ref{fig:medmass_cDvar}. With increasing \cD{} the ground-state energy is lowered in a very regular fashion for all isotopes and for $\cD \approx 4$ we find good agreement with the experimental binding energies over the full mass range. At the same time, the radii are practically independent of \cD{} and in remarkable agreement with experiment in all cases.

We emphasize that there is a clear mismatch between the optimal \cD{} values extracted from few-body systems, medium-mass nuclei, and nuclear-matter saturation. 
Using the $^4$He energy and radius as a guideline (cf. red lines in Fig.~\ref{fig:he4_cDvar}), we would arrive at $\cD\approx 2$ corresponding to the green symbols in Fig.~\ref{fig:medmass_cDvar}. The nuclear-matter studies reported in Ref.~\cite{DrHe19} extract $\cD\approx -3$ from the saturation behaviour for the \NNNLO{} interaction with the same cutoff, but for a regulator with $n=4$. This value leads to a significant underbinding of medium-mass nuclei, as was also shown in Ref.~\cite{HoDr19}. 
Understanding the apparent discrepancy between nuclear matter and medium-mass nuclei will be an important task for future studies.

\paragraph*{Selecting \cD{} in Many-Body Systems.}

We have repeated the above analysis for the \NNLO{} and \NNNLO{} interactions with all three cutoff values and we always find the same basic behavior discussed in Fig.~\ref{fig:medmass_cDvar}. We can select an optimal \cD{} for each chiral order and cutoff, such that the ground-state energy of $^{16}$O is reproduced in simple IM-SRG calculations. Note that we only consider integer values for \cD{} for this selection. Given the limited accuracy of the many-body scheme used in this step, we do not attempt a rigorous fit. The resulting values for the low-energy constants are summarized in Tab.~\ref{tab:lecs}. In addition to the interactions with consistent chiral orders in the NN and 3N sector, denoted by \NNLO{} and \NNNLO{}, we also considered the case of NN interactions at \NNNLO{} combined with 3N interaction at \NNLO{}, denoted by \NNNLO{}'. The optimal \cD{} values show two interesting systematics: (i) they are similar for all different orders with a fixed cutoff, (ii) they are rather large for the smallest cutoff but decrease systematically with increasing cutoff. Table \ref{tab:lecs} also reports the ground-state energy and radius of $^4$He obtained with the respective interactions.

\paragraph*{Medium-Mass Nuclei and Uncertainties.}
Based on this set of interactions we can address the various sources of theory uncertainties. There is already some experience in assessing the uncertainties of the many-body method itself. Various comparisons of different many-body methods for a fixed SRG-evolved Hamiltonian, e.g. in Refs.~\cite{HeBo14,BiLa14,HeBi13,RoBi12}, typically indicated an uncertainty of 1--2\%, e.g., due to the restriction to normal-ordered two-body terms in the IM-SRG formulation. Additional uncertainties due to the free-space SRG evolution and the model space truncations can be shown to be small. Combining all of these effects, we estimate the many-body uncertainties to be on the order of 2\%.

More significant are the uncertainties resulting from the chiral interaction itself. As we have a sequence of interactions from \LO\ to \NNNLO\ for three different cutoff values each, we can explicitly quantify the uncertainties due to the chiral interactions. We use the order-by-order behavior of the observables at fixed cutoff to determine an uncertainty following Refs.~\cite{EpKr15,BiCa16,BiCa18}. The uncertainty of an observable $X_{\NNNLO}$ at order \NNNLO\, e.g., is given by $\max( Q\, | X_{\NNNLO} - X_{\NNLO} |, Q^2\, | X_{\NNLO} - X_{\NLO} |, Q^3\,| X_{\NLO} - X_{\LO} |, Q^5\,| X_{\LO} | )$. 
The expansion parameter $Q$ is estimated by the ratio of a typical momentum scale in the medium-mass regime over the breakdown scale, which results in $Q\approx1/3$. A detailed discussion of these scale estimates can be found in Ref.~\cite{BiCa18}.
The cutoff dependence is used to validate the uncertainty estimate---ideally, we expect the different cutoffs to give results compatible within uncertainties. 
An investigation of more comprehensive schemes, e.g., in a Bayesian framework along the lines of Ref.~\cite{MeWe17,EpGo19b}, will be subject of future work. 

\begin{figure}
\includegraphics[width=1\columnwidth]{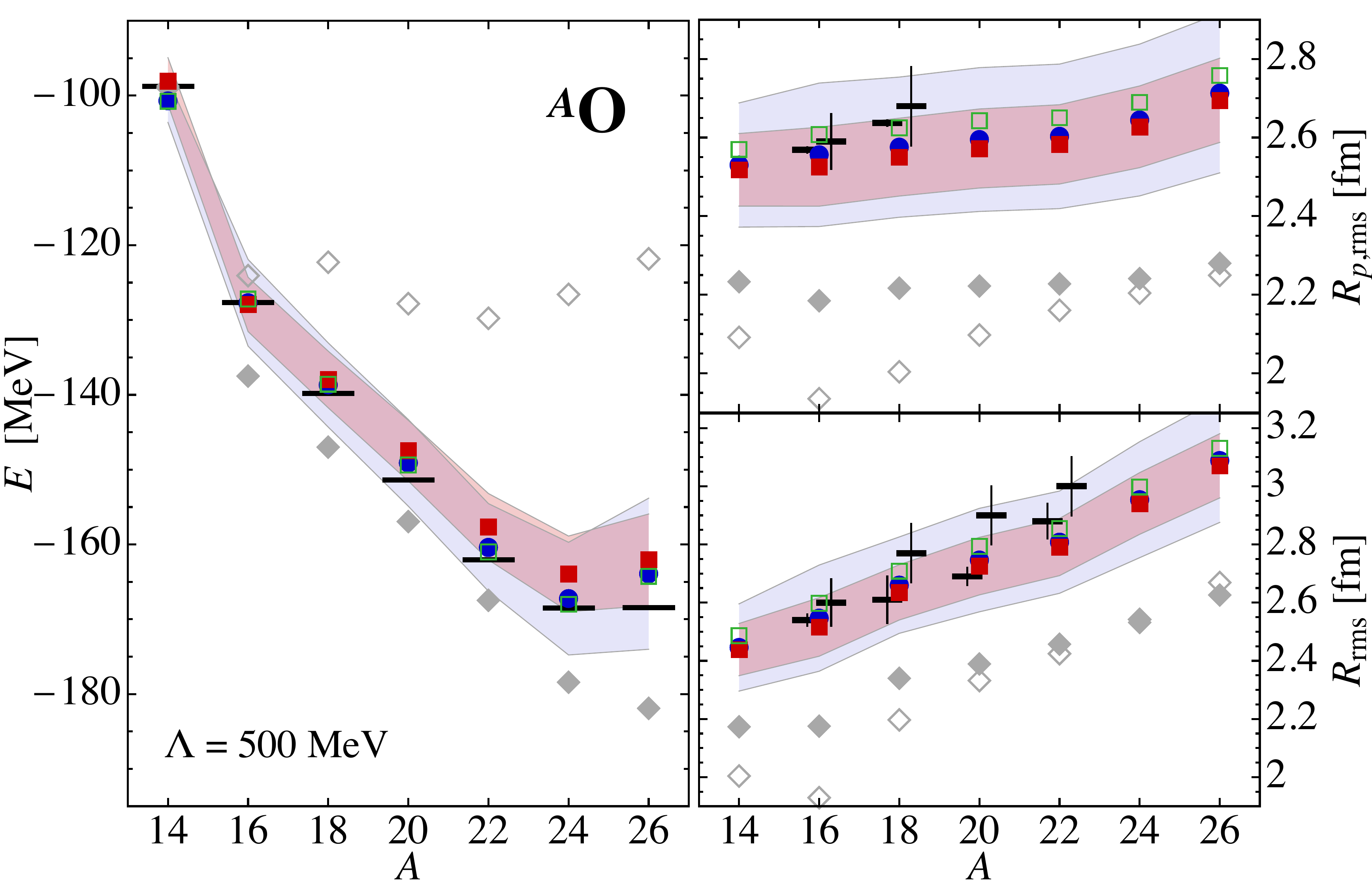}
\caption{Ground-state energies, point-proton rms radii, and mass rms radii of even oxygen isotopes obtained in the IM-NCSM for the \LO{} (open gray diamonds), \NLO{} (solid gray diamonds), \NNLO{} (blue circles), \NNNLO{} (red boxes), and \NNNLO{}' (open green boxes) interactions at $\Lambda=500$\,MeV. Experimental data is indicate by black bars, where two sets of data with error bars are shown for the radii: For proton radii  experimental data is taken from \cite{AnMa13} (left-hand symbols) and \cite{LaSo16} (right-hand symbols), for  mass radii data extracted from interaction cross-sections (left) and from proton scattering (right), discussed in Ref.~\cite{LaSo16}, is shown.\label{fig:oxygen}}
\end{figure}

The ground-states energies and point-proton radii of closed-shell isotopes from oxygen to nickel obtained for the different cutoffs and different chiral orders with uncertainty bands indicating the combined interaction and many-body uncertainties are presented in Fig.~\ref{fig:medmass}. The general picture is remarkable for a number of reasons: (i) the results at \NNLO\ and \NNNLO\ agree extremely well, even without considering the uncertainties; (ii) consequently, the uncertainty bands are nested and generally shrink systematically; (iii) at \NNNLO\ the interaction and many-body uncertainties are comparable, while at \NNLO\ the interaction uncertainties dominate; (iv) results are very stable across the different cutoffs and agree within uncertainties; (v) ground-state energies and point-proton radii agree with experiment within uncertainties for all isotopes considered here.  

The agreement of energies and radii among the different orders and the different cutoffs, and the agreement with experiment, is far from trivial. As we discussed earlier, the majority of existing chiral interactions are not able to reproduce these systematics. As a further cross-check, Fig.~\ref{fig:medmass} also shows the results with the mixed-order \NNNLO' interactions. They also agree with the consistent \NNLO\ and \NNNLO\ interactions within uncertainties, which highlights the robustness of this family of interactions.

\paragraph*{Oxygen Isotopes.}
As an example for applications to open-shell nuclei, we consider the even oxygen isotopes from $^{14}$O to $^{26}$O as shown in Fig.~\ref{fig:oxygen}. For these calculations we use the IM-NCSM with an $N^{\text{ref}}_{\max}=0$ reference state and the same uncertainty quantification protocol as for the medium-mass isotopes including interaction and many-body uncertainties. As before, the ground-state energies and radii at \NNLO{}, \NNNLO{}, and \NNNLO' agree very well with each other and with experiment. The dripline at $^{24}$O is clearly reproduced with all interactions starting from \NNLO{}. We have included both, point-proton and matter rms radii in order to compare to evaluations of the matter rms radii for the neutron-rich oxygen isotopes from Ref.~\cite{LaSo16}. Taking into account the difference between matter radii extracted from interaction cross-sections and proton scattering as well as the experimental and the theory uncertainties, we find good agreement with the available data.

\begin{figure}
\includegraphics[width=0.95\columnwidth]{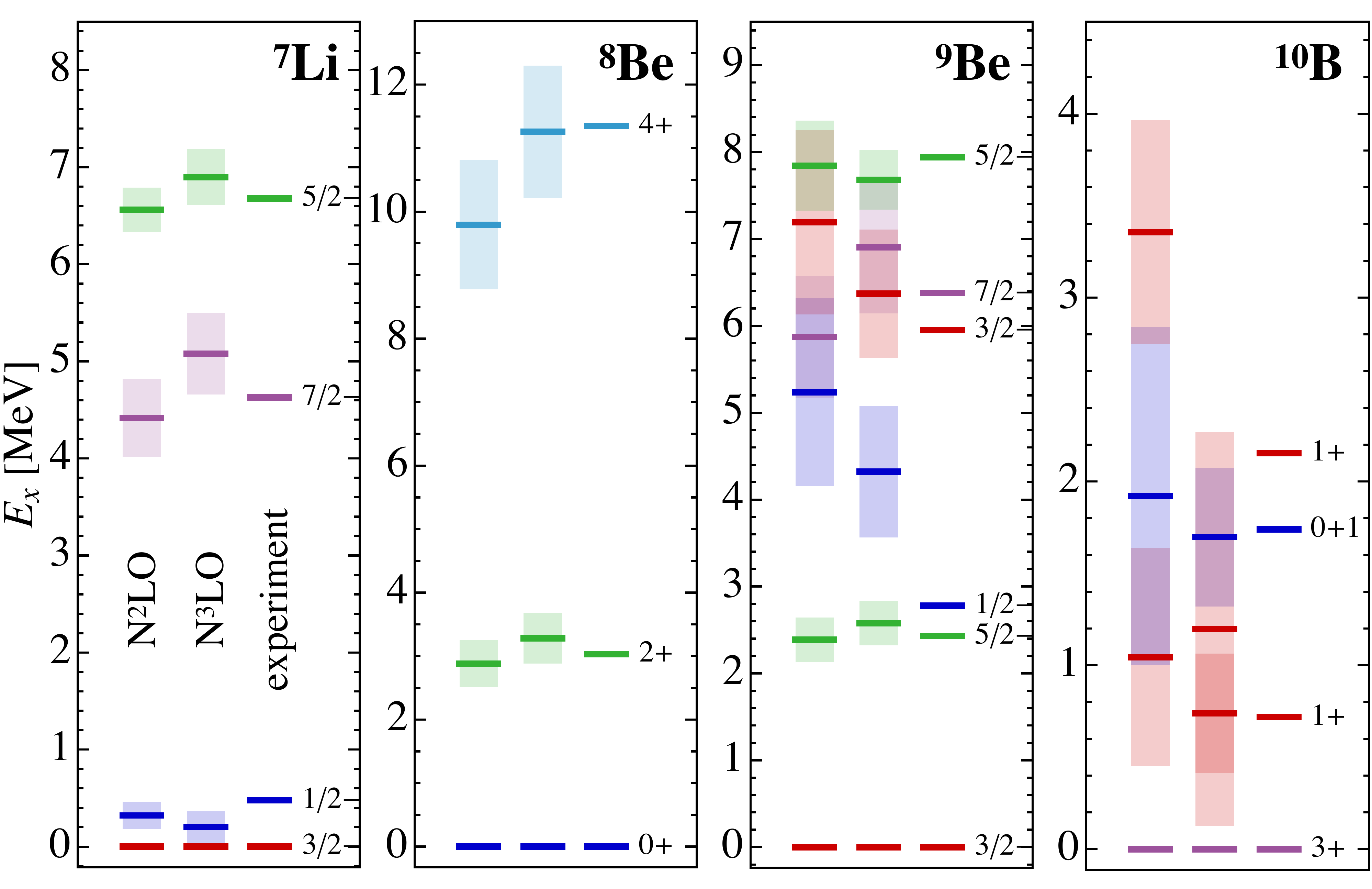}
\caption{Excitation spectra of selected p-shell nuclei from NCSM calculations up to $N_{\max}=10$ for the \NNLO{} and \NNNLO{} interactions at $\Lambda=500$\,MeV. The uncertainty bands show the combined interaction and many-body uncertainties for each state (see text).\label{fig:spectra}}
\end{figure}

\paragraph*{Excitation Spectra.}
Going beyond ground-state observables, Fig.~\ref{fig:spectra} presents the excitation spectra for selected p-shell nuclei obtained in NCSM. We use the order-by-order behaviour of the excitation energies to assess the interaction uncertainties in the same scheme discussed before, the many-body uncertainties are estimated from the difference of results for the two largest values on $N_{\max}$. Generally the spectra agree very well with experiment within uncertainties. One notable exception is the $\tfrac{1}{2}^-$ state in $^9$Be, which appears $1.5$ MeV too high. It was shown in Ref.~\cite{LaNa15} that this state is strongly affected by continuum degrees of freedom, which are not included here. Another interesting case is the second $1^+$ state in $^{10}$B, which appears $1$ MeV too high at \NNLO\ and $1$ MeV too low at \NNNLO\, however, with a very large uncertainties. This state is obviously very sensitive to details of the interaction and shows that spectra and spectroscopy are the obvious next step for validating this new family of interactions.

\paragraph*{Conclusions.}
We have constructed and applied a new family of chiral NN+3N interactions up to \NNNLO{} using non-local regulators for three different cutoff values. They provide an accurate description of two-body phase shifts and, at the same time, reproduce experimental ground-state energies and radii up into the medium-mass regime. In addition, we performed a systematic uncertainty analysis based on the order-by-oder behavior of nuclear observables, which reveals a very robust convergence pattern. This family of interactions will enable a large variety of \emph{ab initio} nuclear structure calculations with fully quantified uncertainties and it promises superior accuracy for many different nuclear observables compared to previous generations of interactions. It also highlights remaining challenges, e.g., the incompatibility of the optimum \cD{} for nuclear matter, medium-mass nuclei, and few-body systems.

\begin{acknowledgments}
\paragraph*{Acknowledgments.}

This work is supported by the Deutsche Forschungsgemeinschaft (DFG, German Research Foundation) Projektnummer 279384907, SFB 1245, the Helmholtz International Center for FAIR (HIC for FAIR), and the BMBF through contracts 05P15RDFN1 and 05P18RDFN1 (NuSTAR.DA). Numerical calculations have been performed on the LICHTENBERG cluster at the computing center of the TU Darmstadt.

\end{acknowledgments}


\begin{thebibliography}{43}%
\makeatletter
\providecommand \@ifxundefined [1]{%
 \@ifx{#1\undefined}
}%
\providecommand \@ifnum [1]{%
 \ifnum #1\expandafter \@firstoftwo
 \else \expandafter \@secondoftwo
 \fi
}%
\providecommand \@ifx [1]{%
 \ifx #1\expandafter \@firstoftwo
 \else \expandafter \@secondoftwo
 \fi
}%
\providecommand \natexlab [1]{#1}%
\providecommand \enquote  [1]{``#1''}%
\providecommand \bibnamefont  [1]{#1}%
\providecommand \bibfnamefont [1]{#1}%
\providecommand \citenamefont [1]{#1}%
\providecommand \href@noop [0]{\@secondoftwo}%
\providecommand \href [0]{\begingroup \@sanitize@url \@href}%
\providecommand \@href[1]{\@@startlink{#1}\@@href}%
\providecommand \@@href[1]{\endgroup#1\@@endlink}%
\providecommand \@sanitize@url [0]{\catcode `\\12\catcode `\$12\catcode
  `\&12\catcode `\#12\catcode `\^12\catcode `\_12\catcode `\%12\relax}%
\providecommand \@@startlink[1]{}%
\providecommand \@@endlink[0]{}%
\providecommand \url  [0]{\begingroup\@sanitize@url \@url }%
\providecommand \@url [1]{\endgroup\@href {#1}{\urlprefix }}%
\providecommand \urlprefix  [0]{URL }%
\providecommand \Eprint [0]{\href }%
\providecommand \doibase [0]{http://dx.doi.org/}%
\providecommand \selectlanguage [0]{\@gobble}%
\providecommand \bibinfo  [0]{\@secondoftwo}%
\providecommand \bibfield  [0]{\@secondoftwo}%
\providecommand \translation [1]{[#1]}%
\providecommand \BibitemOpen [0]{}%
\providecommand \bibitemStop [0]{}%
\providecommand \bibitemNoStop [0]{.\EOS\space}%
\providecommand \EOS [0]{\spacefactor3000\relax}%
\providecommand \BibitemShut  [1]{\csname bibitem#1\endcsname}%
\let\auto@bib@innerbib\@empty
\bibitem [{\citenamefont {Taniuchi}\ \emph {et~al.}(2019)\citenamefont
  {Taniuchi}, \citenamefont {Santamaria}, \citenamefont {Doornenbal},
  \citenamefont {Obertelli}, \citenamefont {Yoneda}, \citenamefont {Authelet},
  \citenamefont {Baba}, \citenamefont {Calvet}, \citenamefont {Chateau},
  \citenamefont {Corsi}, \citenamefont {Delbart}, \citenamefont {Gheller},
  \citenamefont {Gillibert}, \citenamefont {Holt}, \citenamefont {Isobe},
  \citenamefont {Lapoux}, \citenamefont {Matsushita}, \citenamefont {Menendez},
  \citenamefont {Momiyama}, \citenamefont {Motobayashi}, \citenamefont
  {Niikura}, \citenamefont {Nowacki}, \citenamefont {Ogata}, \citenamefont
  {Otsu}, \citenamefont {Otsuka}, \citenamefont {Peron}, \citenamefont {Peru},
  \citenamefont {Peyaud}, \citenamefont {Pollacco}, \citenamefont {Poves},
  \citenamefont {Rousse}, \citenamefont {Sakurai}, \citenamefont {Schwenk},
  \citenamefont {Shiga}, \citenamefont {Simonis}, \citenamefont {Stroberg},
  \citenamefont {Takeuchi}, \citenamefont {Tsunoda}, \citenamefont {Uesaka},
  \citenamefont {Wang}, \citenamefont {Browne}, \citenamefont {Chung},
  \citenamefont {Dombradi}, \citenamefont {Franchoo}, \citenamefont {Giacoppo},
  \citenamefont {Gottardo}, \citenamefont {Hady{\'n}ska-Klek}, \citenamefont
  {Korkulu}, \citenamefont {Koyama}, \citenamefont {Kubota}, \citenamefont
  {Lee}, \citenamefont {Lettmann}, \citenamefont {Louchart}, \citenamefont
  {Lozeva}, \citenamefont {Matsui}, \citenamefont {Miyazaki}, \citenamefont
  {Nishimura}, \citenamefont {Olivier}, \citenamefont {Ota}, \citenamefont
  {Patel}, \citenamefont {Sahin}, \citenamefont {Shand}, \citenamefont
  {Soderstrom}, \citenamefont {Stefan}, \citenamefont {Steppenbeck},
  \citenamefont {Sumikama}, \citenamefont {Suzuki}, \citenamefont {Vajta},
  \citenamefont {Werner}, \citenamefont {Wu},\ and\ \citenamefont
  {Xu}}]{TaSa19}%
  \BibitemOpen
  \bibfield  {author} {\bibinfo {author} {\bibfnamefont {R.}~\bibnamefont
  {Taniuchi}}, \bibinfo {author} {\bibfnamefont {C.}~\bibnamefont
  {Santamaria}}, \bibinfo {author} {\bibfnamefont {P.}~\bibnamefont
  {Doornenbal}}, \bibinfo {author} {\bibfnamefont {A.}~\bibnamefont
  {Obertelli}}, \bibinfo {author} {\bibfnamefont {K.}~\bibnamefont {Yoneda}},
  \bibinfo {author} {\bibfnamefont {G.}~\bibnamefont {Authelet}}, \bibinfo
  {author} {\bibfnamefont {H.}~\bibnamefont {Baba}}, \bibinfo {author}
  {\bibfnamefont {D.}~\bibnamefont {Calvet}}, \bibinfo {author} {\bibfnamefont
  {F.}~\bibnamefont {Chateau}}, \bibinfo {author} {\bibfnamefont
  {A.}~\bibnamefont {Corsi}}, \bibinfo {author} {\bibfnamefont
  {A.}~\bibnamefont {Delbart}}, \bibinfo {author} {\bibfnamefont {J.~M.}\
  \bibnamefont {Gheller}}, \bibinfo {author} {\bibfnamefont {A.}~\bibnamefont
  {Gillibert}}, \bibinfo {author} {\bibfnamefont {J.~D.}\ \bibnamefont {Holt}},
  \bibinfo {author} {\bibfnamefont {T.}~\bibnamefont {Isobe}}, \bibinfo
  {author} {\bibfnamefont {V.}~\bibnamefont {Lapoux}}, \bibinfo {author}
  {\bibfnamefont {M.}~\bibnamefont {Matsushita}}, \bibinfo {author}
  {\bibfnamefont {J.}~\bibnamefont {Menendez}}, \bibinfo {author}
  {\bibfnamefont {S.}~\bibnamefont {Momiyama}}, \bibinfo {author}
  {\bibfnamefont {T.}~\bibnamefont {Motobayashi}}, \bibinfo {author}
  {\bibfnamefont {M.}~\bibnamefont {Niikura}}, \bibinfo {author} {\bibfnamefont
  {F.}~\bibnamefont {Nowacki}}, \bibinfo {author} {\bibfnamefont
  {K.}~\bibnamefont {Ogata}}, \bibinfo {author} {\bibfnamefont
  {H.}~\bibnamefont {Otsu}}, \bibinfo {author} {\bibfnamefont {T.}~\bibnamefont
  {Otsuka}}, \bibinfo {author} {\bibfnamefont {C.}~\bibnamefont {Peron}},
  \bibinfo {author} {\bibfnamefont {S.}~\bibnamefont {Peru}}, \bibinfo {author}
  {\bibfnamefont {A.}~\bibnamefont {Peyaud}}, \bibinfo {author} {\bibfnamefont
  {E.~C.}\ \bibnamefont {Pollacco}}, \bibinfo {author} {\bibfnamefont
  {A.}~\bibnamefont {Poves}}, \bibinfo {author} {\bibfnamefont {J.~Y.}\
  \bibnamefont {Rousse}}, \bibinfo {author} {\bibfnamefont {H.}~\bibnamefont
  {Sakurai}}, \bibinfo {author} {\bibfnamefont {A.}~\bibnamefont {Schwenk}},
  \bibinfo {author} {\bibfnamefont {Y.}~\bibnamefont {Shiga}}, \bibinfo
  {author} {\bibfnamefont {J.}~\bibnamefont {Simonis}}, \bibinfo {author}
  {\bibfnamefont {S.~R.}\ \bibnamefont {Stroberg}}, \bibinfo {author}
  {\bibfnamefont {S.}~\bibnamefont {Takeuchi}}, \bibinfo {author}
  {\bibfnamefont {Y.}~\bibnamefont {Tsunoda}}, \bibinfo {author} {\bibfnamefont
  {T.}~\bibnamefont {Uesaka}}, \bibinfo {author} {\bibfnamefont
  {H.}~\bibnamefont {Wang}}, \bibinfo {author} {\bibfnamefont {F.}~\bibnamefont
  {Browne}}, \bibinfo {author} {\bibfnamefont {L.~X.}\ \bibnamefont {Chung}},
  \bibinfo {author} {\bibfnamefont {Z.}~\bibnamefont {Dombradi}}, \bibinfo
  {author} {\bibfnamefont {S.}~\bibnamefont {Franchoo}}, \bibinfo {author}
  {\bibfnamefont {F.}~\bibnamefont {Giacoppo}}, \bibinfo {author}
  {\bibfnamefont {A.}~\bibnamefont {Gottardo}}, \bibinfo {author}
  {\bibfnamefont {K.}~\bibnamefont {Hady{\'n}ska-Klek}}, \bibinfo {author}
  {\bibfnamefont {Z.}~\bibnamefont {Korkulu}}, \bibinfo {author} {\bibfnamefont
  {S.}~\bibnamefont {Koyama}}, \bibinfo {author} {\bibfnamefont
  {Y.}~\bibnamefont {Kubota}}, \bibinfo {author} {\bibfnamefont
  {J.}~\bibnamefont {Lee}}, \bibinfo {author} {\bibfnamefont {M.}~\bibnamefont
  {Lettmann}}, \bibinfo {author} {\bibfnamefont {C.}~\bibnamefont {Louchart}},
  \bibinfo {author} {\bibfnamefont {R.}~\bibnamefont {Lozeva}}, \bibinfo
  {author} {\bibfnamefont {K.}~\bibnamefont {Matsui}}, \bibinfo {author}
  {\bibfnamefont {T.}~\bibnamefont {Miyazaki}}, \bibinfo {author}
  {\bibfnamefont {S.}~\bibnamefont {Nishimura}}, \bibinfo {author}
  {\bibfnamefont {L.}~\bibnamefont {Olivier}}, \bibinfo {author} {\bibfnamefont
  {S.}~\bibnamefont {Ota}}, \bibinfo {author} {\bibfnamefont {Z.}~\bibnamefont
  {Patel}}, \bibinfo {author} {\bibfnamefont {E.}~\bibnamefont {Sahin}},
  \bibinfo {author} {\bibfnamefont {C.}~\bibnamefont {Shand}}, \bibinfo
  {author} {\bibfnamefont {P.~A.}\ \bibnamefont {Soderstrom}}, \bibinfo
  {author} {\bibfnamefont {I.}~\bibnamefont {Stefan}}, \bibinfo {author}
  {\bibfnamefont {D.}~\bibnamefont {Steppenbeck}}, \bibinfo {author}
  {\bibfnamefont {T.}~\bibnamefont {Sumikama}}, \bibinfo {author}
  {\bibfnamefont {D.}~\bibnamefont {Suzuki}}, \bibinfo {author} {\bibfnamefont
  {Z.}~\bibnamefont {Vajta}}, \bibinfo {author} {\bibfnamefont
  {V.}~\bibnamefont {Werner}}, \bibinfo {author} {\bibfnamefont
  {J.}~\bibnamefont {Wu}}, \ and\ \bibinfo {author} {\bibfnamefont {Z.~Y.}\
  \bibnamefont {Xu}},\ }\href {\doibase 10.1038/s41586-019-1155-x} {\bibfield
  {journal} {\bibinfo  {journal} {Nature}\ }\textbf {\bibinfo {volume} {569}},\
  \bibinfo {pages} {53} (\bibinfo {year} {2019})}\BibitemShut {NoStop}%
\bibitem [{\citenamefont {Maa\ss{}}\ \emph {et~al.}(2019)\citenamefont
  {Maa\ss{}}, \citenamefont {H\"uther}, \citenamefont {K\"onig}, \citenamefont
  {Kr\"amer}, \citenamefont {Krause}, \citenamefont {Lovato}, \citenamefont
  {M\"uller}, \citenamefont {Pachucki}, \citenamefont {Puchalski},
  \citenamefont {Roth}, \citenamefont {S\'anchez}, \citenamefont {Sommer},
  \citenamefont {Wiringa},\ and\ \citenamefont {N\"ortersh\"auser}}]{MaHu19}%
  \BibitemOpen
  \bibfield  {author} {\bibinfo {author} {\bibfnamefont {B.}~\bibnamefont
  {Maa\ss{}}}, \bibinfo {author} {\bibfnamefont {T.}~\bibnamefont {H\"uther}},
  \bibinfo {author} {\bibfnamefont {K.}~\bibnamefont {K\"onig}}, \bibinfo
  {author} {\bibfnamefont {J.}~\bibnamefont {Kr\"amer}}, \bibinfo {author}
  {\bibfnamefont {J.}~\bibnamefont {Krause}}, \bibinfo {author} {\bibfnamefont
  {A.}~\bibnamefont {Lovato}}, \bibinfo {author} {\bibfnamefont
  {P.}~\bibnamefont {M\"uller}}, \bibinfo {author} {\bibfnamefont
  {K.}~\bibnamefont {Pachucki}}, \bibinfo {author} {\bibfnamefont
  {M.}~\bibnamefont {Puchalski}}, \bibinfo {author} {\bibfnamefont
  {R.}~\bibnamefont {Roth}}, \bibinfo {author} {\bibfnamefont {R.}~\bibnamefont
  {S\'anchez}}, \bibinfo {author} {\bibfnamefont {F.}~\bibnamefont {Sommer}},
  \bibinfo {author} {\bibfnamefont {R.~B.}\ \bibnamefont {Wiringa}}, \ and\
  \bibinfo {author} {\bibfnamefont {W.}~\bibnamefont {N\"ortersh\"auser}},\
  }\href {\doibase 10.1103/PhysRevLett.122.182501} {\bibfield  {journal}
  {\bibinfo  {journal} {Phys. Rev. Lett.}\ }\textbf {\bibinfo {volume} {122}},\
  \bibinfo {pages} {182501} (\bibinfo {year} {2019})}\BibitemShut {NoStop}%
\bibitem [{\citenamefont {Gysbers}\ \emph {et~al.}(2019)\citenamefont
  {Gysbers}, \citenamefont {Hagen}, \citenamefont {Holt}, \citenamefont
  {Jansen}, \citenamefont {Morris}, \citenamefont {Navr{\'a}til}, \citenamefont
  {Papenbrock}, \citenamefont {Quaglioni}, \citenamefont {Schwenk},
  \citenamefont {Stroberg},\ and\ \citenamefont {Wendt}}]{GyHa19}%
  \BibitemOpen
  \bibfield  {author} {\bibinfo {author} {\bibfnamefont {P.}~\bibnamefont
  {Gysbers}}, \bibinfo {author} {\bibfnamefont {G.}~\bibnamefont {Hagen}},
  \bibinfo {author} {\bibfnamefont {J.~D.}\ \bibnamefont {Holt}}, \bibinfo
  {author} {\bibfnamefont {G.~R.}\ \bibnamefont {Jansen}}, \bibinfo {author}
  {\bibfnamefont {T.~D.}\ \bibnamefont {Morris}}, \bibinfo {author}
  {\bibfnamefont {P.}~\bibnamefont {Navr{\'a}til}}, \bibinfo {author}
  {\bibfnamefont {T.}~\bibnamefont {Papenbrock}}, \bibinfo {author}
  {\bibfnamefont {S.}~\bibnamefont {Quaglioni}}, \bibinfo {author}
  {\bibfnamefont {A.}~\bibnamefont {Schwenk}}, \bibinfo {author} {\bibfnamefont
  {S.~R.}\ \bibnamefont {Stroberg}}, \ and\ \bibinfo {author} {\bibfnamefont
  {K.~A.}\ \bibnamefont {Wendt}},\ }\href {\doibase 10.1038/s41567-019-0450-7}
  {\bibfield  {journal} {\bibinfo  {journal} {Nature Physics}\ }\textbf
  {\bibinfo {volume} {15}},\ \bibinfo {pages} {428} (\bibinfo {year}
  {2019})}\BibitemShut {NoStop}%
\bibitem [{\citenamefont {Morris}\ \emph {et~al.}(2018)\citenamefont {Morris},
  \citenamefont {Simonis}, \citenamefont {Stroberg}, \citenamefont {Stumpf},
  \citenamefont {Hagen}, \citenamefont {Holt}, \citenamefont {Jansen},
  \citenamefont {Papenbrock}, \citenamefont {Roth},\ and\ \citenamefont
  {Schwenk}}]{MoSi18}%
  \BibitemOpen
  \bibfield  {author} {\bibinfo {author} {\bibfnamefont {T.~D.}\ \bibnamefont
  {Morris}}, \bibinfo {author} {\bibfnamefont {J.}~\bibnamefont {Simonis}},
  \bibinfo {author} {\bibfnamefont {S.~R.}\ \bibnamefont {Stroberg}}, \bibinfo
  {author} {\bibfnamefont {C.}~\bibnamefont {Stumpf}}, \bibinfo {author}
  {\bibfnamefont {G.}~\bibnamefont {Hagen}}, \bibinfo {author} {\bibfnamefont
  {J.~D.}\ \bibnamefont {Holt}}, \bibinfo {author} {\bibfnamefont {G.~R.}\
  \bibnamefont {Jansen}}, \bibinfo {author} {\bibfnamefont {T.}~\bibnamefont
  {Papenbrock}}, \bibinfo {author} {\bibfnamefont {R.}~\bibnamefont {Roth}}, \
  and\ \bibinfo {author} {\bibfnamefont {A.}~\bibnamefont {Schwenk}},\ }\href
  {\doibase 10.1103/PhysRevLett.120.152503} {\bibfield  {journal} {\bibinfo
  {journal} {Phys. Rev. Lett.}\ }\textbf {\bibinfo {volume} {120}},\ \bibinfo
  {pages} {152503} (\bibinfo {year} {2018})}\BibitemShut {NoStop}%
\bibitem [{\citenamefont {Garcia~Ruiz}\ \emph {et~al.}(2016)\citenamefont
  {Garcia~Ruiz}, \citenamefont {Bissell}, \citenamefont {Blaum}, \citenamefont
  {Ekstr{\"o}m}, \citenamefont {Fr{\"o}mmgen}, \citenamefont {Hagen},
  \citenamefont {Hammen}, \citenamefont {Hebeler}, \citenamefont {Holt},
  \citenamefont {Jansen}, \citenamefont {Kowalska}, \citenamefont {Kreim},
  \citenamefont {Nazarewicz}, \citenamefont {Neugart}, \citenamefont {Neyens},
  \citenamefont {N{\"o}rtersh{\"a}user}, \citenamefont {Papenbrock},
  \citenamefont {Papuga}, \citenamefont {Schwenk}, \citenamefont {Simonis},
  \citenamefont {Wendt},\ and\ \citenamefont {Yordanov}}]{GaBi16}%
  \BibitemOpen
  \bibfield  {author} {\bibinfo {author} {\bibfnamefont {R.~F.}\ \bibnamefont
  {Garcia~Ruiz}}, \bibinfo {author} {\bibfnamefont {M.~L.}\ \bibnamefont
  {Bissell}}, \bibinfo {author} {\bibfnamefont {K.}~\bibnamefont {Blaum}},
  \bibinfo {author} {\bibfnamefont {A.}~\bibnamefont {Ekstr{\"o}m}}, \bibinfo
  {author} {\bibfnamefont {N.}~\bibnamefont {Fr{\"o}mmgen}}, \bibinfo {author}
  {\bibfnamefont {G.}~\bibnamefont {Hagen}}, \bibinfo {author} {\bibfnamefont
  {M.}~\bibnamefont {Hammen}}, \bibinfo {author} {\bibfnamefont
  {K.}~\bibnamefont {Hebeler}}, \bibinfo {author} {\bibfnamefont {J.~D.}\
  \bibnamefont {Holt}}, \bibinfo {author} {\bibfnamefont {G.~R.}\ \bibnamefont
  {Jansen}}, \bibinfo {author} {\bibfnamefont {M.}~\bibnamefont {Kowalska}},
  \bibinfo {author} {\bibfnamefont {K.}~\bibnamefont {Kreim}}, \bibinfo
  {author} {\bibfnamefont {W.}~\bibnamefont {Nazarewicz}}, \bibinfo {author}
  {\bibfnamefont {R.}~\bibnamefont {Neugart}}, \bibinfo {author} {\bibfnamefont
  {G.}~\bibnamefont {Neyens}}, \bibinfo {author} {\bibfnamefont
  {W.}~\bibnamefont {N{\"o}rtersh{\"a}user}}, \bibinfo {author} {\bibfnamefont
  {T.}~\bibnamefont {Papenbrock}}, \bibinfo {author} {\bibfnamefont
  {J.}~\bibnamefont {Papuga}}, \bibinfo {author} {\bibfnamefont
  {A.}~\bibnamefont {Schwenk}}, \bibinfo {author} {\bibfnamefont
  {J.}~\bibnamefont {Simonis}}, \bibinfo {author} {\bibfnamefont {K.~A.}\
  \bibnamefont {Wendt}}, \ and\ \bibinfo {author} {\bibfnamefont {D.~T.}\
  \bibnamefont {Yordanov}},\ }\href {https://doi.org/10.1038/nphys3645}
  {\bibfield  {journal} {\bibinfo  {journal} {Nature Physics}\ }\textbf
  {\bibinfo {volume} {12}},\ \bibinfo {pages} {594} (\bibinfo {year}
  {2016})}\BibitemShut {NoStop}%
\bibitem [{\citenamefont {Entem}\ and\ \citenamefont
  {Machleidt}(2003)}]{EnMa03}%
  \BibitemOpen
  \bibfield  {author} {\bibinfo {author} {\bibfnamefont {D.~R.}\ \bibnamefont
  {Entem}}\ and\ \bibinfo {author} {\bibfnamefont {R.}~\bibnamefont
  {Machleidt}},\ }\href@noop {} {\bibfield  {journal} {\bibinfo  {journal}
  {Phys. Rev. C}\ }\textbf {\bibinfo {volume} {68}},\ \bibinfo {pages}
  {041001(R)} (\bibinfo {year} {2003})}\BibitemShut {NoStop}%
\bibitem [{\citenamefont {Navr\'atil}(2007)}]{Na07}%
  \BibitemOpen
  \bibfield  {author} {\bibinfo {author} {\bibfnamefont {P.}~\bibnamefont
  {Navr\'atil}},\ }\href {\doibase 10.1007/s00601-007-0193-3} {\bibfield
  {journal} {\bibinfo  {journal} {Few Body Systems}\ }\textbf {\bibinfo
  {volume} {41}},\ \bibinfo {pages} {117} (\bibinfo {year} {2007})}\BibitemShut
  {NoStop}%
\bibitem [{\citenamefont {Gazit}\ \emph {et~al.}(2009)\citenamefont {Gazit},
  \citenamefont {Quaglioni},\ and\ \citenamefont {Navr\'atil}}]{GaQu09}%
  \BibitemOpen
  \bibfield  {author} {\bibinfo {author} {\bibfnamefont {D.}~\bibnamefont
  {Gazit}}, \bibinfo {author} {\bibfnamefont {S.}~\bibnamefont {Quaglioni}}, \
  and\ \bibinfo {author} {\bibfnamefont {P.}~\bibnamefont {Navr\'atil}},\
  }\href {\doibase 10.1103/PhysRevLett.103.102502} {\bibfield  {journal}
  {\bibinfo  {journal} {Phys. Rev. Lett.}\ }\textbf {\bibinfo {volume} {103}},\
  \bibinfo {pages} {102502} (\bibinfo {year} {2009})}\BibitemShut {NoStop}%
\bibitem [{\citenamefont {Gazit}\ \emph {et~al.}(2019)\citenamefont {Gazit},
  \citenamefont {Quaglioni},\ and\ \citenamefont {Navr\'atil}}]{GaQu19}%
  \BibitemOpen
  \bibfield  {author} {\bibinfo {author} {\bibfnamefont {D.}~\bibnamefont
  {Gazit}}, \bibinfo {author} {\bibfnamefont {S.}~\bibnamefont {Quaglioni}}, \
  and\ \bibinfo {author} {\bibfnamefont {P.}~\bibnamefont {Navr\'atil}},\
  }\href {\doibase 10.1103/PhysRevLett.122.029901} {\bibfield  {journal}
  {\bibinfo  {journal} {Phys. Rev. Lett.}\ }\textbf {\bibinfo {volume} {122}},\
  \bibinfo {pages} {029901(E)} (\bibinfo {year} {2019})}\BibitemShut {NoStop}%
\bibitem [{\citenamefont {Epelbaum}\ \emph {et~al.}(2002)\citenamefont
  {Epelbaum}, \citenamefont {Nogga}, \citenamefont {Gl\"ockle}, \citenamefont
  {Kamada}, \citenamefont {Mei\ss{}ner},\ and\ \citenamefont
  {Witala}}]{EpNo02}%
  \BibitemOpen
  \bibfield  {author} {\bibinfo {author} {\bibfnamefont {E.}~\bibnamefont
  {Epelbaum}}, \bibinfo {author} {\bibfnamefont {A.}~\bibnamefont {Nogga}},
  \bibinfo {author} {\bibfnamefont {W.}~\bibnamefont {Gl\"ockle}}, \bibinfo
  {author} {\bibfnamefont {H.}~\bibnamefont {Kamada}}, \bibinfo {author}
  {\bibfnamefont {U.-G.}\ \bibnamefont {Mei\ss{}ner}}, \ and\ \bibinfo {author}
  {\bibfnamefont {H.}~\bibnamefont {Witala}},\ }\href@noop {} {\bibfield
  {journal} {\bibinfo  {journal} {Phys. Rev. C}\ }\textbf {\bibinfo {volume}
  {66}},\ \bibinfo {pages} {064001} (\bibinfo {year} {2002})}\BibitemShut
  {NoStop}%
\bibitem [{\citenamefont {Epelbaum}\ \emph {et~al.}(2005)\citenamefont
  {Epelbaum}, \citenamefont {Gl\"ockle},\ and\ \citenamefont
  {Mei\ss{}ner}}]{EpGl05}%
  \BibitemOpen
  \bibfield  {author} {\bibinfo {author} {\bibfnamefont {E.}~\bibnamefont
  {Epelbaum}}, \bibinfo {author} {\bibfnamefont {W.}~\bibnamefont {Gl\"ockle}},
  \ and\ \bibinfo {author} {\bibfnamefont {U.-G.}\ \bibnamefont
  {Mei\ss{}ner}},\ }\href
  {http://www.sciencedirect.com/science/article/pii/S0375947404010747}
  {\bibfield  {journal} {\bibinfo  {journal} {Nucl. Phys. A}\ }\textbf
  {\bibinfo {volume} {747}},\ \bibinfo {pages} {362} (\bibinfo {year}
  {2005})}\BibitemShut {NoStop}%
\bibitem [{\citenamefont {Calci}\ and\ \citenamefont {Roth}(2016)}]{CaRo16}%
  \BibitemOpen
  \bibfield  {author} {\bibinfo {author} {\bibfnamefont {A.}~\bibnamefont
  {Calci}}\ and\ \bibinfo {author} {\bibfnamefont {R.}~\bibnamefont {Roth}},\
  }\href {\doibase 10.1103/PhysRevC.94.014322} {\bibfield  {journal} {\bibinfo
  {journal} {Phys. Rev. C}\ }\textbf {\bibinfo {volume} {94}},\ \bibinfo
  {pages} {014322} (\bibinfo {year} {2016})}\BibitemShut {NoStop}%
\bibitem [{\citenamefont {Lapoux}\ \emph {et~al.}(2016)\citenamefont {Lapoux},
  \citenamefont {Som\`a}, \citenamefont {Barbieri}, \citenamefont {Hergert},
  \citenamefont {Holt},\ and\ \citenamefont {Stroberg}}]{LaSo16}%
  \BibitemOpen
  \bibfield  {author} {\bibinfo {author} {\bibfnamefont {V.}~\bibnamefont
  {Lapoux}}, \bibinfo {author} {\bibfnamefont {V.}~\bibnamefont {Som\`a}},
  \bibinfo {author} {\bibfnamefont {C.}~\bibnamefont {Barbieri}}, \bibinfo
  {author} {\bibfnamefont {H.}~\bibnamefont {Hergert}}, \bibinfo {author}
  {\bibfnamefont {J.~D.}\ \bibnamefont {Holt}}, \ and\ \bibinfo {author}
  {\bibfnamefont {S.~R.}\ \bibnamefont {Stroberg}},\ }\href {\doibase
  10.1103/PhysRevLett.117.052501} {\bibfield  {journal} {\bibinfo  {journal}
  {Phys. Rev. Lett.}\ }\textbf {\bibinfo {volume} {117}},\ \bibinfo {pages}
  {052501} (\bibinfo {year} {2016})}\BibitemShut {NoStop}%
\bibitem [{\citenamefont {Binder}\ \emph {et~al.}(2014)\citenamefont {Binder},
  \citenamefont {Langhammer}, \citenamefont {Calci},\ and\ \citenamefont
  {Roth}}]{BiLa14}%
  \BibitemOpen
  \bibfield  {author} {\bibinfo {author} {\bibfnamefont {S.}~\bibnamefont
  {Binder}}, \bibinfo {author} {\bibfnamefont {J.}~\bibnamefont {Langhammer}},
  \bibinfo {author} {\bibfnamefont {A.}~\bibnamefont {Calci}}, \ and\ \bibinfo
  {author} {\bibfnamefont {R.}~\bibnamefont {Roth}},\ }\href@noop {} {\bibfield
   {journal} {\bibinfo  {journal} {Phys. Lett. B}\ }\textbf {\bibinfo {volume}
  {736}},\ \bibinfo {pages} {119} (\bibinfo {year} {2014})}\BibitemShut
  {NoStop}%
\bibitem [{\citenamefont {Ekstr\"om}\ \emph {et~al.}(2015)\citenamefont
  {Ekstr\"om}, \citenamefont {Jansen}, \citenamefont {Wendt}, \citenamefont
  {Hagen}, \citenamefont {Papenbrock}, \citenamefont {Carlsson}, \citenamefont
  {Forss\'en}, \citenamefont {Hjorth-Jensen}, \citenamefont {Navr\'atil},\ and\
  \citenamefont {Nazarewicz}}]{EkHa15}%
  \BibitemOpen
  \bibfield  {author} {\bibinfo {author} {\bibfnamefont {A.}~\bibnamefont
  {Ekstr\"om}}, \bibinfo {author} {\bibfnamefont {G.~R.}\ \bibnamefont
  {Jansen}}, \bibinfo {author} {\bibfnamefont {K.~A.}\ \bibnamefont {Wendt}},
  \bibinfo {author} {\bibfnamefont {G.}~\bibnamefont {Hagen}}, \bibinfo
  {author} {\bibfnamefont {T.}~\bibnamefont {Papenbrock}}, \bibinfo {author}
  {\bibfnamefont {B.~D.}\ \bibnamefont {Carlsson}}, \bibinfo {author}
  {\bibfnamefont {C.}~\bibnamefont {Forss\'en}}, \bibinfo {author}
  {\bibfnamefont {M.}~\bibnamefont {Hjorth-Jensen}}, \bibinfo {author}
  {\bibfnamefont {P.}~\bibnamefont {Navr\'atil}}, \ and\ \bibinfo {author}
  {\bibfnamefont {W.}~\bibnamefont {Nazarewicz}},\ }\href {\doibase
  10.1103/PhysRevC.91.051301} {\bibfield  {journal} {\bibinfo  {journal} {Phys.
  Rev. C}\ }\textbf {\bibinfo {volume} {91}},\ \bibinfo {pages} {051301(R)}
  (\bibinfo {year} {2015})}\BibitemShut {NoStop}%
\bibitem [{\citenamefont {Hebeler}\ \emph {et~al.}(2011)\citenamefont
  {Hebeler}, \citenamefont {Bogner}, \citenamefont {Furnstahl}, \citenamefont
  {Nogga},\ and\ \citenamefont {Schwenk}}]{HeBo11}%
  \BibitemOpen
  \bibfield  {author} {\bibinfo {author} {\bibfnamefont {K.}~\bibnamefont
  {Hebeler}}, \bibinfo {author} {\bibfnamefont {S.~K.}\ \bibnamefont {Bogner}},
  \bibinfo {author} {\bibfnamefont {R.~J.}\ \bibnamefont {Furnstahl}}, \bibinfo
  {author} {\bibfnamefont {A.}~\bibnamefont {Nogga}}, \ and\ \bibinfo {author}
  {\bibfnamefont {A.}~\bibnamefont {Schwenk}},\ }\href@noop {} {\bibfield
  {journal} {\bibinfo  {journal} {Phys. Rev. C}\ }\textbf {\bibinfo {volume}
  {83}},\ \bibinfo {pages} {031301(R)} (\bibinfo {year} {2011})}\BibitemShut
  {NoStop}%
\bibitem [{\citenamefont {Som{\`a}}\ \emph {et~al.}(2019)\citenamefont
  {Som{\`a}}, \citenamefont {Navr{\'a}til}, \citenamefont {Raimondi},
  \citenamefont {Barbieri},\ and\ \citenamefont {Duguet}}]{SoNa19}%
  \BibitemOpen
  \bibfield  {author} {\bibinfo {author} {\bibfnamefont {V.}~\bibnamefont
  {Som{\`a}}}, \bibinfo {author} {\bibfnamefont {P.}~\bibnamefont
  {Navr{\'a}til}}, \bibinfo {author} {\bibfnamefont {F.}~\bibnamefont
  {Raimondi}}, \bibinfo {author} {\bibfnamefont {C.}~\bibnamefont {Barbieri}},
  \ and\ \bibinfo {author} {\bibfnamefont {T.}~\bibnamefont {Duguet}},\
  }\href@noop {} {\enquote {\bibinfo {title} {Novel chiral hamiltonian and
  observables in light and medium-mass nuclei},}\ } (\bibinfo {year} {2019}),\
  \Eprint {http://arxiv.org/abs/1907.09790} {arXiv:1907.09790 [nucl-th]}
  \BibitemShut {NoStop}%
\bibitem [{\citenamefont {Epelbaum}\ \emph {et~al.}(2015)\citenamefont
  {Epelbaum}, \citenamefont {Krebs},\ and\ \citenamefont
  {Mei{\ss}ner}}]{EpKr15}%
  \BibitemOpen
  \bibfield  {author} {\bibinfo {author} {\bibfnamefont {E.}~\bibnamefont
  {Epelbaum}}, \bibinfo {author} {\bibfnamefont {H.}~\bibnamefont {Krebs}}, \
  and\ \bibinfo {author} {\bibfnamefont {U.~G.}\ \bibnamefont {Mei{\ss}ner}},\
  }\href {\doibase 10.1140/epja/i2015-15053-8} {\bibfield  {journal} {\bibinfo
  {journal} {Eur. Phys. J. A}\ }\textbf {\bibinfo {volume} {51}},\ \bibinfo
  {pages} {53} (\bibinfo {year} {2015})}\BibitemShut {NoStop}%
\bibitem [{\citenamefont {Binder}\ \emph {et~al.}(2016)\citenamefont {Binder},
  \citenamefont {Calci}, \citenamefont {Epelbaum}, \citenamefont {Furnstahl},
  \citenamefont {Golak}, \citenamefont {Hebeler}, \citenamefont {Kamada},
  \citenamefont {Krebs}, \citenamefont {Langhammer}, \citenamefont {Liebig},
  \citenamefont {Maris}, \citenamefont {Mei\ss{}ner}, \citenamefont {Minossi},
  \citenamefont {Nogga}, \citenamefont {Potter}, \citenamefont {Roth},
  \citenamefont {Skibi\ifmmode~\acute{n}\else \'{n}\fi{}ski}, \citenamefont
  {Topolnicki}, \citenamefont {Vary},\ and\ \citenamefont
  {Wita\l{}a}}]{BiCa16}%
  \BibitemOpen
  \bibfield  {author} {\bibinfo {author} {\bibfnamefont {S.}~\bibnamefont
  {Binder}}, \bibinfo {author} {\bibfnamefont {A.}~\bibnamefont {Calci}},
  \bibinfo {author} {\bibfnamefont {E.}~\bibnamefont {Epelbaum}}, \bibinfo
  {author} {\bibfnamefont {R.~J.}\ \bibnamefont {Furnstahl}}, \bibinfo {author}
  {\bibfnamefont {J.}~\bibnamefont {Golak}}, \bibinfo {author} {\bibfnamefont
  {K.}~\bibnamefont {Hebeler}}, \bibinfo {author} {\bibfnamefont
  {H.}~\bibnamefont {Kamada}}, \bibinfo {author} {\bibfnamefont
  {H.}~\bibnamefont {Krebs}}, \bibinfo {author} {\bibfnamefont
  {J.}~\bibnamefont {Langhammer}}, \bibinfo {author} {\bibfnamefont
  {S.}~\bibnamefont {Liebig}}, \bibinfo {author} {\bibfnamefont
  {P.}~\bibnamefont {Maris}}, \bibinfo {author} {\bibfnamefont {U.-G.}\
  \bibnamefont {Mei\ss{}ner}}, \bibinfo {author} {\bibfnamefont
  {D.}~\bibnamefont {Minossi}}, \bibinfo {author} {\bibfnamefont
  {A.}~\bibnamefont {Nogga}}, \bibinfo {author} {\bibfnamefont
  {H.}~\bibnamefont {Potter}}, \bibinfo {author} {\bibfnamefont
  {R.}~\bibnamefont {Roth}}, \bibinfo {author} {\bibfnamefont {R.}~\bibnamefont
  {Skibi\ifmmode~\acute{n}\else \'{n}\fi{}ski}}, \bibinfo {author}
  {\bibfnamefont {K.}~\bibnamefont {Topolnicki}}, \bibinfo {author}
  {\bibfnamefont {J.~P.}\ \bibnamefont {Vary}}, \ and\ \bibinfo {author}
  {\bibfnamefont {H.}~\bibnamefont {Wita\l{}a}} (\bibinfo {collaboration}
  {LENPIC Collaboration}),\ }\href {\doibase 10.1103/PhysRevC.93.044002}
  {\bibfield  {journal} {\bibinfo  {journal} {Phys. Rev. C}\ }\textbf {\bibinfo
  {volume} {93}},\ \bibinfo {pages} {044002} (\bibinfo {year}
  {2016})}\BibitemShut {NoStop}%
\bibitem [{\citenamefont {Binder}\ \emph {et~al.}(2018)\citenamefont {Binder},
  \citenamefont {Calci}, \citenamefont {Epelbaum}, \citenamefont {Furnstahl},
  \citenamefont {Golak}, \citenamefont {Hebeler}, \citenamefont {H\"uther},
  \citenamefont {Kamada}, \citenamefont {Krebs}, \citenamefont {Maris},
  \citenamefont {Mei\ss{}ner}, \citenamefont {Nogga}, \citenamefont {Roth},
  \citenamefont {Skibi\ifmmode~\acute{n}\else \'{n}\fi{}ski}, \citenamefont
  {Topolnicki}, \citenamefont {Vary}, \citenamefont {Vobig},\ and\
  \citenamefont {Wita\l{}a}}]{BiCa18}%
  \BibitemOpen
  \bibfield  {author} {\bibinfo {author} {\bibfnamefont {S.}~\bibnamefont
  {Binder}}, \bibinfo {author} {\bibfnamefont {A.}~\bibnamefont {Calci}},
  \bibinfo {author} {\bibfnamefont {E.}~\bibnamefont {Epelbaum}}, \bibinfo
  {author} {\bibfnamefont {R.~J.}\ \bibnamefont {Furnstahl}}, \bibinfo {author}
  {\bibfnamefont {J.}~\bibnamefont {Golak}}, \bibinfo {author} {\bibfnamefont
  {K.}~\bibnamefont {Hebeler}}, \bibinfo {author} {\bibfnamefont
  {T.}~\bibnamefont {H\"uther}}, \bibinfo {author} {\bibfnamefont
  {H.}~\bibnamefont {Kamada}}, \bibinfo {author} {\bibfnamefont
  {H.}~\bibnamefont {Krebs}}, \bibinfo {author} {\bibfnamefont
  {P.}~\bibnamefont {Maris}}, \bibinfo {author} {\bibfnamefont {U.-G.}\
  \bibnamefont {Mei\ss{}ner}}, \bibinfo {author} {\bibfnamefont
  {A.}~\bibnamefont {Nogga}}, \bibinfo {author} {\bibfnamefont
  {R.}~\bibnamefont {Roth}}, \bibinfo {author} {\bibfnamefont {R.}~\bibnamefont
  {Skibi\ifmmode~\acute{n}\else \'{n}\fi{}ski}}, \bibinfo {author}
  {\bibfnamefont {K.}~\bibnamefont {Topolnicki}}, \bibinfo {author}
  {\bibfnamefont {J.~P.}\ \bibnamefont {Vary}}, \bibinfo {author}
  {\bibfnamefont {K.}~\bibnamefont {Vobig}}, \ and\ \bibinfo {author}
  {\bibfnamefont {H.}~\bibnamefont {Wita\l{}a}} (\bibinfo {collaboration}
  {LENPIC Collaboration}),\ }\href {\doibase 10.1103/PhysRevC.98.014002}
  {\bibfield  {journal} {\bibinfo  {journal} {Phys. Rev. C}\ }\textbf {\bibinfo
  {volume} {98}},\ \bibinfo {pages} {014002} (\bibinfo {year}
  {2018})}\BibitemShut {NoStop}%
\bibitem [{\citenamefont {Drischler}\ \emph {et~al.}(2019)\citenamefont
  {Drischler}, \citenamefont {Hebeler},\ and\ \citenamefont
  {Schwenk}}]{DrHe19}%
  \BibitemOpen
  \bibfield  {author} {\bibinfo {author} {\bibfnamefont {C.}~\bibnamefont
  {Drischler}}, \bibinfo {author} {\bibfnamefont {K.}~\bibnamefont {Hebeler}},
  \ and\ \bibinfo {author} {\bibfnamefont {A.}~\bibnamefont {Schwenk}},\ }\href
  {\doibase 10.1103/PhysRevLett.122.042501} {\bibfield  {journal} {\bibinfo
  {journal} {Phys. Rev. Lett.}\ }\textbf {\bibinfo {volume} {122}},\ \bibinfo
  {pages} {042501} (\bibinfo {year} {2019})}\BibitemShut {NoStop}%
\bibitem [{\citenamefont {Epelbaum}\ \emph
  {et~al.}(2019{\natexlab{a}})\citenamefont {Epelbaum}, \citenamefont {Golak},
  \citenamefont {Hebeler}, \citenamefont {H\"uther}, \citenamefont {Kamada},
  \citenamefont {Krebs}, \citenamefont {Maris}, \citenamefont {Mei\ss{}ner},
  \citenamefont {Nogga}, \citenamefont {Roth}, \citenamefont
  {Skibi\ifmmode~\acute{n}\else \'{n}\fi{}ski}, \citenamefont {Topolnicki},
  \citenamefont {Vary}, \citenamefont {Vobig},\ and\ \citenamefont
  {Wita\l{}a}}]{EpGo19}%
  \BibitemOpen
  \bibfield  {author} {\bibinfo {author} {\bibfnamefont {E.}~\bibnamefont
  {Epelbaum}}, \bibinfo {author} {\bibfnamefont {J.}~\bibnamefont {Golak}},
  \bibinfo {author} {\bibfnamefont {K.}~\bibnamefont {Hebeler}}, \bibinfo
  {author} {\bibfnamefont {T.}~\bibnamefont {H\"uther}}, \bibinfo {author}
  {\bibfnamefont {H.}~\bibnamefont {Kamada}}, \bibinfo {author} {\bibfnamefont
  {H.}~\bibnamefont {Krebs}}, \bibinfo {author} {\bibfnamefont
  {P.}~\bibnamefont {Maris}}, \bibinfo {author} {\bibfnamefont {U.-G.}\
  \bibnamefont {Mei\ss{}ner}}, \bibinfo {author} {\bibfnamefont
  {A.}~\bibnamefont {Nogga}}, \bibinfo {author} {\bibfnamefont
  {R.}~\bibnamefont {Roth}}, \bibinfo {author} {\bibfnamefont {R.}~\bibnamefont
  {Skibi\ifmmode~\acute{n}\else \'{n}\fi{}ski}}, \bibinfo {author}
  {\bibfnamefont {K.}~\bibnamefont {Topolnicki}}, \bibinfo {author}
  {\bibfnamefont {J.~P.}\ \bibnamefont {Vary}}, \bibinfo {author}
  {\bibfnamefont {K.}~\bibnamefont {Vobig}}, \ and\ \bibinfo {author}
  {\bibfnamefont {H.}~\bibnamefont {Wita\l{}a}} (\bibinfo {collaboration}
  {LENPIC Collaboration}),\ }\href {\doibase 10.1103/PhysRevC.99.024313}
  {\bibfield  {journal} {\bibinfo  {journal} {Phys. Rev. C}\ }\textbf {\bibinfo
  {volume} {99}},\ \bibinfo {pages} {024313} (\bibinfo {year}
  {2019}{\natexlab{a}})}\BibitemShut {NoStop}%
\bibitem [{\citenamefont {Entem}\ \emph {et~al.}(2017)\citenamefont {Entem},
  \citenamefont {Machleidt},\ and\ \citenamefont {Nosyk}}]{EnMa17}%
  \BibitemOpen
  \bibfield  {author} {\bibinfo {author} {\bibfnamefont {D.~R.}\ \bibnamefont
  {Entem}}, \bibinfo {author} {\bibfnamefont {R.}~\bibnamefont {Machleidt}}, \
  and\ \bibinfo {author} {\bibfnamefont {Y.}~\bibnamefont {Nosyk}},\ }\href
  {\doibase 10.1103/PhysRevC.96.024004} {\bibfield  {journal} {\bibinfo
  {journal} {Phys. Rev. C}\ }\textbf {\bibinfo {volume} {96}},\ \bibinfo
  {pages} {024004} (\bibinfo {year} {2017})}\BibitemShut {NoStop}%
\bibitem [{\citenamefont {Hoppe}\ \emph {et~al.}(2019)\citenamefont {Hoppe},
  \citenamefont {Drischler}, \citenamefont {Hebeler}, \citenamefont {Schwenk},\
  and\ \citenamefont {Simonis}}]{HoDr19}%
  \BibitemOpen
  \bibfield  {author} {\bibinfo {author} {\bibfnamefont {J.}~\bibnamefont
  {Hoppe}}, \bibinfo {author} {\bibfnamefont {C.}~\bibnamefont {Drischler}},
  \bibinfo {author} {\bibfnamefont {K.}~\bibnamefont {Hebeler}}, \bibinfo
  {author} {\bibfnamefont {A.}~\bibnamefont {Schwenk}}, \ and\ \bibinfo
  {author} {\bibfnamefont {J.}~\bibnamefont {Simonis}},\ }\href {\doibase
  10.1103/PhysRevC.100.024318} {\bibfield  {journal} {\bibinfo  {journal}
  {Phys. Rev. C}\ }\textbf {\bibinfo {volume} {100}},\ \bibinfo {pages}
  {024318} (\bibinfo {year} {2019})}\BibitemShut {NoStop}%
\bibitem [{\citenamefont {Hergert}(2016)}]{He16}%
  \BibitemOpen
  \bibfield  {author} {\bibinfo {author} {\bibfnamefont {H.}~\bibnamefont
  {Hergert}},\ }\href {\doibase 10.1088/1402-4896/92/2/023002} {\bibfield
  {journal} {\bibinfo  {journal} {Physica Scripta}\ }\textbf {\bibinfo {volume}
  {92}},\ \bibinfo {pages} {023002} (\bibinfo {year} {2016})}\BibitemShut
  {NoStop}%
\bibitem [{\citenamefont {Hergert}\ \emph {et~al.}(2016)\citenamefont
  {Hergert}, \citenamefont {Bogner}, \citenamefont {Morris}, \citenamefont
  {Schwenk},\ and\ \citenamefont {Tsukiyama}}]{HeBo16}%
  \BibitemOpen
  \bibfield  {author} {\bibinfo {author} {\bibfnamefont {H.}~\bibnamefont
  {Hergert}}, \bibinfo {author} {\bibfnamefont {S.}~\bibnamefont {Bogner}},
  \bibinfo {author} {\bibfnamefont {T.}~\bibnamefont {Morris}}, \bibinfo
  {author} {\bibfnamefont {A.}~\bibnamefont {Schwenk}}, \ and\ \bibinfo
  {author} {\bibfnamefont {K.}~\bibnamefont {Tsukiyama}},\ }\href {\doibase
  https://doi.org/10.1016/j.physrep.2015.12.007} {\bibfield  {journal}
  {\bibinfo  {journal} {Phys. Rep.}\ }\textbf {\bibinfo {volume} {621}},\
  \bibinfo {pages} {165} (\bibinfo {year} {2016})}\BibitemShut {NoStop}%
\bibitem [{\citenamefont {Hergert}\ \emph
  {et~al.}(2013{\natexlab{a}})\citenamefont {Hergert}, \citenamefont {Bogner},
  \citenamefont {Binder}, \citenamefont {Calci}, \citenamefont {Langhammer},
  \citenamefont {Roth},\ and\ \citenamefont {Schwenk}}]{HeBo13}%
  \BibitemOpen
  \bibfield  {author} {\bibinfo {author} {\bibfnamefont {H.}~\bibnamefont
  {Hergert}}, \bibinfo {author} {\bibfnamefont {S.~K.}\ \bibnamefont {Bogner}},
  \bibinfo {author} {\bibfnamefont {S.}~\bibnamefont {Binder}}, \bibinfo
  {author} {\bibfnamefont {A.}~\bibnamefont {Calci}}, \bibinfo {author}
  {\bibfnamefont {J.}~\bibnamefont {Langhammer}}, \bibinfo {author}
  {\bibfnamefont {R.}~\bibnamefont {Roth}}, \ and\ \bibinfo {author}
  {\bibfnamefont {A.}~\bibnamefont {Schwenk}},\ }\href@noop {} {\bibfield
  {journal} {\bibinfo  {journal} {Phys. Rev. C}\ }\textbf {\bibinfo {volume}
  {87}},\ \bibinfo {pages} {034307} (\bibinfo {year}
  {2013}{\natexlab{a}})}\BibitemShut {NoStop}%
\bibitem [{\citenamefont {Tsukiyama}\ \emph {et~al.}(2011)\citenamefont
  {Tsukiyama}, \citenamefont {Bogner},\ and\ \citenamefont {Schwenk}}]{TsBo11}%
  \BibitemOpen
  \bibfield  {author} {\bibinfo {author} {\bibfnamefont {K.}~\bibnamefont
  {Tsukiyama}}, \bibinfo {author} {\bibfnamefont {S.~K.}\ \bibnamefont
  {Bogner}}, \ and\ \bibinfo {author} {\bibfnamefont {A.}~\bibnamefont
  {Schwenk}},\ }\href {\doibase 10.1103/PhysRevLett.106.222502} {\bibfield
  {journal} {\bibinfo  {journal} {Phys. Rev. Lett.}\ }\textbf {\bibinfo
  {volume} {106}},\ \bibinfo {pages} {222502} (\bibinfo {year}
  {2011})}\BibitemShut {NoStop}%
\bibitem [{\citenamefont {Roth}\ \emph {et~al.}(2011)\citenamefont {Roth},
  \citenamefont {Langhammer}, \citenamefont {Calci}, \citenamefont {Binder},\
  and\ \citenamefont {Navr{\'a}til}}]{RoLa11}%
  \BibitemOpen
  \bibfield  {author} {\bibinfo {author} {\bibfnamefont {R.}~\bibnamefont
  {Roth}}, \bibinfo {author} {\bibfnamefont {J.}~\bibnamefont {Langhammer}},
  \bibinfo {author} {\bibfnamefont {A.}~\bibnamefont {Calci}}, \bibinfo
  {author} {\bibfnamefont {S.}~\bibnamefont {Binder}}, \ and\ \bibinfo {author}
  {\bibfnamefont {P.}~\bibnamefont {Navr{\'a}til}},\ }\href {\doibase
  10.1103/PhysRevLett.107.072501} {\bibfield  {journal} {\bibinfo  {journal}
  {Phys. Rev. Lett.}\ }\textbf {\bibinfo {volume} {107}},\ \bibinfo {pages}
  {072501} (\bibinfo {year} {2011})}\BibitemShut {NoStop}%
\bibitem [{\citenamefont {Roth}\ \emph {et~al.}(2014)\citenamefont {Roth},
  \citenamefont {Calci}, \citenamefont {Langhammer},\ and\ \citenamefont
  {Binder}}]{RoCa14}%
  \BibitemOpen
  \bibfield  {author} {\bibinfo {author} {\bibfnamefont {R.}~\bibnamefont
  {Roth}}, \bibinfo {author} {\bibfnamefont {A.}~\bibnamefont {Calci}},
  \bibinfo {author} {\bibfnamefont {J.}~\bibnamefont {Langhammer}}, \ and\
  \bibinfo {author} {\bibfnamefont {S.}~\bibnamefont {Binder}},\ }\href
  {\doibase 10.1103/PhysRevC.90.024325} {\bibfield  {journal} {\bibinfo
  {journal} {Phys. Rev. C}\ }\textbf {\bibinfo {volume} {90}},\ \bibinfo
  {pages} {024325} (\bibinfo {year} {2014})}\BibitemShut {NoStop}%
\bibitem [{\citenamefont {Tichai}\ \emph {et~al.}(2019)\citenamefont {Tichai},
  \citenamefont {M\"uller}, \citenamefont {Vobig},\ and\ \citenamefont
  {Roth}}]{TiMu19}%
  \BibitemOpen
  \bibfield  {author} {\bibinfo {author} {\bibfnamefont {A.}~\bibnamefont
  {Tichai}}, \bibinfo {author} {\bibfnamefont {J.}~\bibnamefont {M\"uller}},
  \bibinfo {author} {\bibfnamefont {K.}~\bibnamefont {Vobig}}, \ and\ \bibinfo
  {author} {\bibfnamefont {R.}~\bibnamefont {Roth}},\ }\href {\doibase
  10.1103/PhysRevC.99.034321} {\bibfield  {journal} {\bibinfo  {journal} {Phys.
  Rev. C}\ }\textbf {\bibinfo {volume} {99}},\ \bibinfo {pages} {034321}
  (\bibinfo {year} {2019})}\BibitemShut {NoStop}%
\bibitem [{\citenamefont {Gebrerufael}\ \emph {et~al.}(2017)\citenamefont
  {Gebrerufael}, \citenamefont {Vobig}, \citenamefont {Hergert},\ and\
  \citenamefont {Roth}}]{GeVo17}%
  \BibitemOpen
  \bibfield  {author} {\bibinfo {author} {\bibfnamefont {E.}~\bibnamefont
  {Gebrerufael}}, \bibinfo {author} {\bibfnamefont {K.}~\bibnamefont {Vobig}},
  \bibinfo {author} {\bibfnamefont {H.}~\bibnamefont {Hergert}}, \ and\
  \bibinfo {author} {\bibfnamefont {R.}~\bibnamefont {Roth}},\ }\href {\doibase
  10.1103/PhysRevLett.118.152503} {\bibfield  {journal} {\bibinfo  {journal}
  {Phys. Rev. Lett.}\ }\textbf {\bibinfo {volume} {118}},\ \bibinfo {pages}
  {152503} (\bibinfo {year} {2017})}\BibitemShut {NoStop}%
\bibitem [{\citenamefont {Hebeler}\ \emph {et~al.}(2015)\citenamefont
  {Hebeler}, \citenamefont {Krebs}, \citenamefont {Epelbaum}, \citenamefont
  {Golak},\ and\ \citenamefont {Skibi\ifmmode~\acute{n}\else
  \'{n}\fi{}ski}}]{HeKr15}%
  \BibitemOpen
  \bibfield  {author} {\bibinfo {author} {\bibfnamefont {K.}~\bibnamefont
  {Hebeler}}, \bibinfo {author} {\bibfnamefont {H.}~\bibnamefont {Krebs}},
  \bibinfo {author} {\bibfnamefont {E.}~\bibnamefont {Epelbaum}}, \bibinfo
  {author} {\bibfnamefont {J.}~\bibnamefont {Golak}}, \ and\ \bibinfo {author}
  {\bibfnamefont {R.}~\bibnamefont {Skibi\ifmmode~\acute{n}\else
  \'{n}\fi{}ski}},\ }\href {\doibase 10.1103/PhysRevC.91.044001} {\bibfield
  {journal} {\bibinfo  {journal} {Phys. Rev. C}\ }\textbf {\bibinfo {volume}
  {91}},\ \bibinfo {pages} {044001} (\bibinfo {year} {2015})}\BibitemShut
  {NoStop}%
\bibitem [{\citenamefont {Platter}\ \emph {et~al.}(2005)\citenamefont
  {Platter}, \citenamefont {Hammer},\ and\ \citenamefont
  {Mei{\ss}ner}}]{PlHa05}%
  \BibitemOpen
  \bibfield  {author} {\bibinfo {author} {\bibfnamefont {L.}~\bibnamefont
  {Platter}}, \bibinfo {author} {\bibfnamefont {H.-W.}\ \bibnamefont {Hammer}},
  \ and\ \bibinfo {author} {\bibfnamefont {U.-G.}\ \bibnamefont
  {Mei{\ss}ner}},\ }\href {\doibase
  https://doi.org/10.1016/j.physletb.2004.12.068} {\bibfield  {journal}
  {\bibinfo  {journal} {Phys. Lett. B}\ }\textbf {\bibinfo {volume} {607}},\
  \bibinfo {pages} {254} (\bibinfo {year} {2005})}\BibitemShut {NoStop}%
\bibitem [{\citenamefont {Nogga}\ \emph {et~al.}(2000)\citenamefont {Nogga},
  \citenamefont {Kamada},\ and\ \citenamefont {Gl\"ockle}}]{NoKa00}%
  \BibitemOpen
  \bibfield  {author} {\bibinfo {author} {\bibfnamefont {A.}~\bibnamefont
  {Nogga}}, \bibinfo {author} {\bibfnamefont {H.}~\bibnamefont {Kamada}}, \
  and\ \bibinfo {author} {\bibfnamefont {W.}~\bibnamefont {Gl\"ockle}},\
  }\href@noop {} {\bibfield  {journal} {\bibinfo  {journal} {Phys. Rev. Lett.}\
  }\textbf {\bibinfo {volume} {85}},\ \bibinfo {pages} {944} (\bibinfo {year}
  {2000})}\BibitemShut {NoStop}%
\bibitem [{\citenamefont {Wang}\ \emph {et~al.}(2017)\citenamefont {Wang},
  \citenamefont {Audi}, \citenamefont {Kondev}, \citenamefont {Huang},
  \citenamefont {Naimi},\ and\ \citenamefont {Xu}}]{AME2016}%
  \BibitemOpen
  \bibfield  {author} {\bibinfo {author} {\bibfnamefont {M.}~\bibnamefont
  {Wang}}, \bibinfo {author} {\bibfnamefont {G.}~\bibnamefont {Audi}}, \bibinfo
  {author} {\bibfnamefont {F.~G.}\ \bibnamefont {Kondev}}, \bibinfo {author}
  {\bibfnamefont {W.}~\bibnamefont {Huang}}, \bibinfo {author} {\bibfnamefont
  {S.}~\bibnamefont {Naimi}}, \ and\ \bibinfo {author} {\bibfnamefont
  {X.}~\bibnamefont {Xu}},\ }\href {\doibase 10.1088/1674-1137/41/3/030003}
  {\bibfield  {journal} {\bibinfo  {journal} {Chinese Physics C}\ }\textbf
  {\bibinfo {volume} {41}},\ \bibinfo {pages} {030003} (\bibinfo {year}
  {2017})}\BibitemShut {NoStop}%
\bibitem [{\citenamefont {Angeli}\ and\ \citenamefont
  {Marinova}(2013)}]{AnMa13}%
  \BibitemOpen
  \bibfield  {author} {\bibinfo {author} {\bibfnamefont {I.}~\bibnamefont
  {Angeli}}\ and\ \bibinfo {author} {\bibfnamefont {K.}~\bibnamefont
  {Marinova}},\ }\href {\doibase https://doi.org/10.1016/j.adt.2011.12.006}
  {\bibfield  {journal} {\bibinfo  {journal} {Atomic Data and Nuclear Data
  Tables}\ }\textbf {\bibinfo {volume} {99}},\ \bibinfo {pages} {69} (\bibinfo
  {year} {2013})}\BibitemShut {NoStop}%
\bibitem [{\citenamefont {Hergert}\ \emph {et~al.}(2014)\citenamefont
  {Hergert}, \citenamefont {Bogner}, \citenamefont {Morris}, \citenamefont
  {Binder}, \citenamefont {Calci}, \citenamefont {Langhammer},\ and\
  \citenamefont {Roth}}]{HeBo14}%
  \BibitemOpen
  \bibfield  {author} {\bibinfo {author} {\bibfnamefont {H.}~\bibnamefont
  {Hergert}}, \bibinfo {author} {\bibfnamefont {S.~K.}\ \bibnamefont {Bogner}},
  \bibinfo {author} {\bibfnamefont {T.~D.}\ \bibnamefont {Morris}}, \bibinfo
  {author} {\bibfnamefont {S.}~\bibnamefont {Binder}}, \bibinfo {author}
  {\bibfnamefont {A.}~\bibnamefont {Calci}}, \bibinfo {author} {\bibfnamefont
  {J.}~\bibnamefont {Langhammer}}, \ and\ \bibinfo {author} {\bibfnamefont
  {R.}~\bibnamefont {Roth}},\ }\href {\doibase 10.1103/PhysRevC.90.041302}
  {\bibfield  {journal} {\bibinfo  {journal} {Phys. Rev. C}\ }\textbf {\bibinfo
  {volume} {90}},\ \bibinfo {pages} {041302(R)} (\bibinfo {year}
  {2014})}\BibitemShut {NoStop}%
\bibitem [{\citenamefont {Hergert}\ \emph
  {et~al.}(2013{\natexlab{b}})\citenamefont {Hergert}, \citenamefont {Binder},
  \citenamefont {Calci}, \citenamefont {Langhammer},\ and\ \citenamefont
  {Roth}}]{HeBi13}%
  \BibitemOpen
  \bibfield  {author} {\bibinfo {author} {\bibfnamefont {H.}~\bibnamefont
  {Hergert}}, \bibinfo {author} {\bibfnamefont {S.}~\bibnamefont {Binder}},
  \bibinfo {author} {\bibfnamefont {A.}~\bibnamefont {Calci}}, \bibinfo
  {author} {\bibfnamefont {J.}~\bibnamefont {Langhammer}}, \ and\ \bibinfo
  {author} {\bibfnamefont {R.}~\bibnamefont {Roth}},\ }\href@noop {} {\bibfield
   {journal} {\bibinfo  {journal} {Phys. Rev. Lett.}\ }\textbf {\bibinfo
  {volume} {110}},\ \bibinfo {pages} {242501} (\bibinfo {year}
  {2013}{\natexlab{b}})}\BibitemShut {NoStop}%
\bibitem [{\citenamefont {Roth}\ \emph {et~al.}(2012)\citenamefont {Roth},
  \citenamefont {Binder}, \citenamefont {Vobig}, \citenamefont {Calci},
  \citenamefont {Langhammer},\ and\ \citenamefont {Navr\'atil}}]{RoBi12}%
  \BibitemOpen
  \bibfield  {author} {\bibinfo {author} {\bibfnamefont {R.}~\bibnamefont
  {Roth}}, \bibinfo {author} {\bibfnamefont {S.}~\bibnamefont {Binder}},
  \bibinfo {author} {\bibfnamefont {K.}~\bibnamefont {Vobig}}, \bibinfo
  {author} {\bibfnamefont {A.}~\bibnamefont {Calci}}, \bibinfo {author}
  {\bibfnamefont {J.}~\bibnamefont {Langhammer}}, \ and\ \bibinfo {author}
  {\bibfnamefont {P.}~\bibnamefont {Navr\'atil}},\ }\href@noop {} {\bibfield
  {journal} {\bibinfo  {journal} {Phys. Rev. Lett.}\ }\textbf {\bibinfo
  {volume} {109}},\ \bibinfo {pages} {052501} (\bibinfo {year}
  {2012})}\BibitemShut {NoStop}%
\bibitem [{\citenamefont {Melendez}\ \emph {et~al.}(2017)\citenamefont
  {Melendez}, \citenamefont {Wesolowski},\ and\ \citenamefont
  {Furnstahl}}]{MeWe17}%
  \BibitemOpen
  \bibfield  {author} {\bibinfo {author} {\bibfnamefont {J.~A.}\ \bibnamefont
  {Melendez}}, \bibinfo {author} {\bibfnamefont {S.}~\bibnamefont
  {Wesolowski}}, \ and\ \bibinfo {author} {\bibfnamefont {R.~J.}\ \bibnamefont
  {Furnstahl}},\ }\href {\doibase 10.1103/PhysRevC.96.024003} {\bibfield
  {journal} {\bibinfo  {journal} {Phys. Rev. C}\ }\textbf {\bibinfo {volume}
  {96}},\ \bibinfo {pages} {024003} (\bibinfo {year} {2017})}\BibitemShut
  {NoStop}%
\bibitem [{\citenamefont {Epelbaum}\ \emph
  {et~al.}(2019{\natexlab{b}})\citenamefont {Epelbaum}, \citenamefont {Golak},
  \citenamefont {Hebeler}, \citenamefont {Kamada}, \citenamefont {Krebs},
  \citenamefont {Mei{\ss}ner}, \citenamefont {Nogga}, \citenamefont {Reinert},
  \citenamefont {Skibi{\'n}ski}, \citenamefont {Topolnicki}, \citenamefont
  {Volkotrub},\ and\ \citenamefont {Wita{\l}a}}]{EpGo19b}%
  \BibitemOpen
  \bibfield  {author} {\bibinfo {author} {\bibfnamefont {E.}~\bibnamefont
  {Epelbaum}}, \bibinfo {author} {\bibfnamefont {J.}~\bibnamefont {Golak}},
  \bibinfo {author} {\bibfnamefont {K.}~\bibnamefont {Hebeler}}, \bibinfo
  {author} {\bibfnamefont {H.}~\bibnamefont {Kamada}}, \bibinfo {author}
  {\bibfnamefont {H.}~\bibnamefont {Krebs}}, \bibinfo {author} {\bibfnamefont
  {U.~G.}\ \bibnamefont {Mei{\ss}ner}}, \bibinfo {author} {\bibfnamefont
  {A.}~\bibnamefont {Nogga}}, \bibinfo {author} {\bibfnamefont
  {P.}~\bibnamefont {Reinert}}, \bibinfo {author} {\bibfnamefont
  {R.}~\bibnamefont {Skibi{\'n}ski}}, \bibinfo {author} {\bibfnamefont
  {K.}~\bibnamefont {Topolnicki}}, \bibinfo {author} {\bibfnamefont
  {Y.}~\bibnamefont {Volkotrub}}, \ and\ \bibinfo {author} {\bibfnamefont
  {H.}~\bibnamefont {Wita{\l}a}},\ }\href@noop {} {\enquote {\bibinfo {title}
  {Towards high-order calculations of three-nucleon scattering in chiral
  effective field theory},}\ } (\bibinfo {year} {2019}{\natexlab{b}}),\ \Eprint
  {http://arxiv.org/abs/1907.03608} {arXiv:1907.03608 [nucl-th]} \BibitemShut
  {NoStop}%
\bibitem [{\citenamefont {Langhammer}\ \emph {et~al.}(2015)\citenamefont
  {Langhammer}, \citenamefont {Navr\'atil}, \citenamefont {Quaglioni},
  \citenamefont {Hupin}, \citenamefont {Calci},\ and\ \citenamefont
  {Roth}}]{LaNa15}%
  \BibitemOpen
  \bibfield  {author} {\bibinfo {author} {\bibfnamefont {J.}~\bibnamefont
  {Langhammer}}, \bibinfo {author} {\bibfnamefont {P.}~\bibnamefont
  {Navr\'atil}}, \bibinfo {author} {\bibfnamefont {S.}~\bibnamefont
  {Quaglioni}}, \bibinfo {author} {\bibfnamefont {G.}~\bibnamefont {Hupin}},
  \bibinfo {author} {\bibfnamefont {A.}~\bibnamefont {Calci}}, \ and\ \bibinfo
  {author} {\bibfnamefont {R.}~\bibnamefont {Roth}},\ }\href {\doibase
  10.1103/PhysRevC.91.021301} {\bibfield  {journal} {\bibinfo  {journal} {Phys.
  Rev. C}\ }\textbf {\bibinfo {volume} {91}},\ \bibinfo {pages} {021301(R)}
  (\bibinfo {year} {2015})}\BibitemShut {NoStop}%
\end{thebibliography}
\end{document}